\documentclass[aps,prd,reprint,
groupedaddress,nofootinbib,a4paper]{revtex4-2}

\usepackage{amsmath}
\usepackage{amssymb}
\usepackage{graphicx}
\usepackage{lmodern}
\usepackage[utf8]{inputenc}
\usepackage[T1]{fontenc}
\usepackage{microtype}
\usepackage[colorlinks=true, allcolors=blue]{hyperref}
\usepackage[dvipsnames]{xcolor}
\usepackage[normalem]{ulem}
\usepackage{multirow}

\begin{document}

\title{Large black-hole scalar charges induced by cosmology
in Horndeski theories}

\author{Eugeny Babichev}
\email[]{babichev@ijclab.in2p3.fr}
\affiliation{Universit\'e Paris-Saclay, CNRS/IN2P3,
IJCLab, 91405, Orsay, France}

\author{Gilles Esposito-Far\`ese}
\email[]{gef@iap.fr}
\affiliation{${\mathcal{G}}{\mathbb{R}}
\varepsilon{\mathbb{C}}{\mathcal{O}}$,
Institut d'Astrophysique de Paris,
CNRS and Sorbonne Universit\'e, UMR 7095,
98bis boulevard Arago, F-75014 Paris, France}

\author{Ignacy Sawicki}
\email[]{sawicki@fzu.cz}
\affiliation{CEICO, Institute of Physics of the Czech Academy of
Sciences, Na Slovance 1999/2, 182 00, Prague 8, Czechia}

\author{Leonardo G.~Trombetta}
\email[]{trombetta@fzu.cz}
\affiliation{CEICO, Institute of Physics of the Czech Academy of
Sciences, Na Slovance 1999/2, 182 00, Prague 8, Czechia}

\date{\today}

\begin{abstract}
The regularity of black hole solutions, embedded in an expanding
Universe, is studied in a subclass of Horndeski theories, namely
the sum of the simplest quadratic, cubic and quintic actions. We
find that in presence of a time derivative of the scalar field,
driven by the cosmological expansion, this regularity generically
imposes large scalar charges for black holes, even when assuming
strictly no direct coupling of matter to the scalar field. Such
charges cause a significant accretion of the scalar field by the
black holes, driving its local time derivative to a small value.
This phenomenon, together with the Vainshtein screening typical
of these theories, strongly suppresses observable scalar effects.
We show that this full class of models is consistent with
LIGO/Virgo detections of gravitational waves, but that the LISA
mission should be able to constrain the coefficient of the
quintic term at the $10^{-30}$ level in a self-acceleration
scenario, an improvement by 16 orders of magnitude with respect
to what is imposed by the speed of gravitational waves.
\end{abstract}

\maketitle

\section{Introduction}
\label{Sec1}

The uniqueness of black holes is one of the key predictions of
general relativity (GR). The detection of gravitational waves
(GW) from the final stages of the inspiral of binary compact
objects has finally opened up a direct avenue for verifying this
property.

Scalar-tensor theories provide concrete alternative models
allowing for the exploration of consistent compact-object
phenomenology differing from the GR setup. They allow for the
existence of hair attached to scalar charges carried by compact
objects, providing an opportunity for testing gravity.
Static hairy solutions for black holes were first considered
in the context of conformally coupled theories of gravity,
where they are singular on the
horizon~\cite{Bocharova:1970skc,Bekenstein:1974sf}. Neutron
stars, on the other hand, admit regular hairy solutions with a
mass-dependent sensitivity~\cite{Damour:1993hw}, allowing for the
production of dipolar emission from binaries. Observations of
such binary pulsars over a long time span still provide the best
constraints for such non-minimal
couplings~\cite{Freire:2012mg,Kramer:2021jcw}.

More general theories, all part of the Horndeski
class~\cite{Horndeski:1974wa} giving standard second-order
equations of motion, allow for alternative (derivative)
interaction terms with gravity. In the presence of potentials for
the scalar field, these tend to be suppressed and the
phenomenology is similar to the conformal case. When the
scalar-field action is shift symmetric, the new operators can be
important, but alternative no-hair theorems
exist~\cite{Hui:2012qt}. The emission of dipolar radiation does
not take place and therefore these models are not strongly
constrained~\cite{Barausse:2015wia}.

However, the above conclusions are predicated on \emph{both} a
static configuration for the scalar field and asymptotically
trivial boundary conditions at infinity. Already in
Ref.~\cite{Jacobson:1999vr} it was pointed out that an evolving
scalar field at the boundary leads to an induced scalar charge
for a black hole even in conformally coupled theories. This
effect is of utmost relevance for black holes embedded in
cosmology, although for conformally coupled theories the charge
remains hard to observe~\cite{Horbatsch:2011ye}. Secondly, the
existence of a cosmological horizon requires that a static
profile for the scalar field sourced by a charge be augmented by
a time-dependent term~\cite{Glavan:2021adm}, the lack of which
can even lead to singularities on horizons in some
models~\cite{Babichev:2024txe}. The consistent setup for a black
hole in cosmology therefore has \emph{both} an induced scalar
charge and a time-dependent profile. Whether this is a matter of
principle or actually relevant to phenomenology is the central
question of this paper.

Indeed, the fast-rolling scalar field is exploited in cosmology
to provide a mechanism for late-time acceleration different to
the cosmological constant or scalar-field potential
energy~\cite{ArmendarizPicon:2000dh,Deffayet:2010qz}.
Such models can modify the speed of GWs in cosmology
and therefore have been constrained by its
measurement~\cite{Baker:2017hug,Creminelli:2017sry,
Ezquiaga:2017ekz,Sakstein:2017xjx}, nonetheless a wide range of
parameters are still allowed. The typical construction is one
where the canonical kinetic term for the scalar is chosen to have
the wrong sign and therefore fluctuations around the trivial
scalar background are ghosts. The cosmological dynamics then
chooses a non-trivial Lorentz-violating background which
nonetheless asymptotically looks similar to de Sitter spacetime.
This introduces another barrier to the existence of static BH
solutions, since one would then have to connect the spacelike
gradients of the scalar field near the BH to the timelike
gradients at large distances without probing the unstable
backgrounds. A configuration of the scalar gradient that is
time-like everywhere but stationary owing to the shift symmetry
is a natural resolution of this tension.

The fact that the cosmological evolution of $\varphi$ can induce
an effective coupling to a local source has been already
discussed in the literature in different contexts. In particular,
it was noticed in Ref.~\cite{Babichev:2012re} that the
matter-scalar coupling strength is modified due to the
cosmological evolution of $\varphi$, in the case of the cubic
Galileon model. This effect can be attributed to the kinetic
scalar-graviton mixing in presence of material sources, such as
stars. In case of black holes in the background of a
time-dependent scalar field, a non-trivial configuration of the
scalar field also arises, as it has been shown for asymptotic
Minkowski spacetime~\cite{Babichev:2006vx,Babichev:2010kj}.
However, this is owing to a different physical reason in this
case, namely, this is a consequence of the requirement of
non-divergence of observable quantities at the BH horizon.
Therefore a non-trivial scalar configuration arises around a
black hole even in the case of minimal coupling. In fact, the
appearance of a nontrivial scalar configuration around a black
hole for non-zero time derivative $\dot{\varphi}$ can be viewed
as a process of accretion, which is well understood for perfect
fluids, see e.g.~\cite{Babichev:2004yx,Babichev:2013vji}.

A construction of smooth everywhere non-singular solutions,
interpolating between the black hole and cosmological horizons,
may be problematic due to various issues. In
Ref.~\cite{Babichev:2024txe} it was demonstrated that a
particular model involving the linear scalar-Gauss-Bonnet (sGB)
term leads to singular behavior at a horizon. In this paper,
we focus on theories which do not present this issue (see
Appendix~\ref{AppendixA}), i.e., a solution can be constructed
such that observables are regular at both the cosmological and
the black hole horizons. Nonetheless, we demonstrate that another
problem generically arises, related to a smooth transition from
one branch of the solution to another as the distance from the
black hole increases. Making a parallel with accretion of a
perfect fluid, one has to build an analogue of a transsonic
solution, which implies jumping from one branch to another. As we
show in the present paper, it turns out to be challenging to have
smooth branch transitions, when taking into account that normally
two such points arise ---~one in the vicinity of the black-hole
horizon, one at cosmological distances.

The paper is organized as follows. Section~\ref{Sec2} defines the
subclass of Horndeski theories we consider, derives their test
scalar-field solution, and discusses two generic phenomena which
significantly reduce observable scalar-field deviations from
general relativity: Scalar-field accretion by BHs makes their
scalar charge decrease with time, and the Vainshtein mechanism
screens scalar exchanges between two BHs as well as the binary's
energy loss via scalar waves. Section~\ref{Sec4} focuses on the
quadratic plus cubic Galileon model. It shows that the
discriminant of a quadratic equation needs to have double roots
at some precise locations, explains our technique to compute
them, and derives the BH's scalar charge and the local time
derivative of the scalar field needed to get a linear
time-dependent solution in the whole Universe. Section~\ref{Sec5}
considers another subclass of models, involving the simplest
quintic Horndeski action in addition to the standard quadratic
kinetic term for the scalar field. It illustrates the strong
accretion of the scalar field which occurs when too large scalar
charges are predicted for BHs. Section~\ref{Sec6} is devoted to
the phenomenologically richer case of the quadratic, cubic and
quintic kinetic terms together. It is particularly interesting
when cosmology is dominated by the quadratic and cubic Galileon
terms, while the local physics of BHs is dominated by the quintic
term. Section~\ref{Observations} discusses the observational
consequences of these three models in binary black-hole
coalescences. While all of them are consistent with the present
LIGO/Virgo data, we show that the coefficient of the quintic term
will be tightly constrained by LISA. Our conclusions are given in
Section~\ref{Sec7}.

\section{Action and test scalar-field solution}
\label{Sec2}
\subsection{Quadratic, cubic and quintic Horndeski}
\label{Sec2A}
In the present paper, we focus on the quadratic, cubic and
quintic Horndeski actions, together with the standard
Einstein-Hilbert term and a possible cosmological constant. The
corresponding action may be written as
\begin{eqnarray}
S &=& M_\text{Pl}^2 \int\sqrt{-g}\,d^4 x\biggl\{
\frac{R}{2} - \Lambda_\text{bare}
\nonumber\\
&&+G_2(X)
+G_3(X) \Box\varphi
+G_5(X) G^{\mu\nu} \varphi_{\mu\nu}
\nonumber\\
&&-\frac{1}{3} G_5'(X)
\Bigl[\left(\Box \varphi\right)^3
- 3\, \Box\varphi\, \varphi_{\mu\nu}\varphi^{\mu\nu}
+ 2\, \varphi_{\mu\nu}\varphi^{\nu\rho}
\varphi_\rho^{\hphantom{\rho}\mu}\Bigr]
\biggr\}
\nonumber\\
&&+S_\text{matter}[\psi, e^{2\alpha\varphi}g_{\mu\nu}],
\label{eqAction}
\end{eqnarray}
where $M_\text{Pl} \equiv (8\pi G)^{-1/2}$ is the reduced Planck
mass (in units such that $\hbar = c = 1$), $R$ is the curvature
scalar of the metric $g_{\mu\nu}$ with the sign conventions of
Ref.~\cite{Misner:1973prb} (notably the mostly-plus signature),
$\varphi$ is a dimensionless scalar field whose first derivative
is denoted as $\varphi_\mu \equiv \partial_\mu\varphi$, similarly
$\varphi_{\mu\nu} \equiv \nabla_\nu\nabla_\mu\varphi$ for its
second covariant derivative, and $X \equiv - \varphi_\mu^2
= - g^{\mu\nu}\partial_\mu\varphi\partial_\nu\varphi$.

We have also included the action specifying how matter fields,
globally denoted as $\psi$, are universally coupled to a physical
metric which may differ from $g_{\mu\nu}$. The exponential factor
we wrote here means that matter is assumed to be linearly coupled
to the scalar field with a bare coupling constant $\alpha$ (which
is dimensionless). However, its actual coupling generically
differs from $\alpha$ because of the nonlinearities of these
theories~\cite{Babichev:2012re,Anson:2020fum,Sakstein:2014isa,
Kobayashi:2014ida,Babichev:2016jom}. Moreover, we will mainly
focus on black holes in the present work, i.e., vacuum solutions
for which this bare $\alpha$ does not play any role. We will
actually derive that a black hole effective coupling constant to
the scalar field, say $\alpha_\text{BH}$, depends on the
parameters entering the functions $G_2(X)$, $G_3(X)$ and
$G_5(X)$, on the Hubble constant $H$, and on the Schwarzschild
radius $r_S$ of the black hole.

In absence of matter sources,
action~(\ref{eqAction}) only depends on the derivatives of the
scalar field, therefore it is ``shift-symmetric'', i.e.,
invariant when one adds a constant to the scalar field.
Noether's theorem implies that there exists a conserved current
related to this symmetry. This is simply
\begin{equation}
J^\mu \equiv -\frac{1}{\sqrt{-g}}\,\frac{\delta
S}{\delta\partial_\mu \varphi}, \label{eq:phi-eom}
\end{equation}
and its covariant conservation $\nabla_\mu J^\mu = 0$ is the
scalar field equation of motion.

We further restrict action~(\ref{eqAction}) by imposing that
the three functions $G_2(X)$, $G_3(X)$ and $G_5(X)$ are linear
in~$X$:
\begin{subequations}
\label{G235}
\begin{eqnarray}
G_2(X) &=& k_2 X,\label{G2}\\
G_3(X) &=& \frac{k_3}{M^2} X,\label{G3}\\
G_5(X) &=& \frac{k_5}{M^4} X,\label{G5}
\end{eqnarray}
\end{subequations}
where $k_2$, $k_3$ and $k_5$ are dimensionless constants and $M$
is a constant mass parameter. Note that, while these choices of
$G_2(X)$ and $G_3(X)$ correspond to the respective covariant
Galileons, our $G_5(X)$ does not.\footnote{The quintic Galileon
would correspond instead to $G_5(X) \propto X^2$.} Nevertheless,
we will loosely refer to our models as Galileons. Some factors
could easily be reabsorbed in the definition of $\varphi$ and
these constants, but it is useful to keep track of the origin of
the various terms in our results below. Moreover, a positive
value of $k_2$ corresponds to a standard positive-energy degree
of freedom on a background with vanishing scalar derivatives.
Instead, we will take a negative value of $k_2$ ---~this
implies that the standard configuration with vanishing scalar
derivatives is unstable. However, there now exists a well-behaved
attractor in which both the background and fluctuations of the
scalar have positive energy, which generates an accelerated
expansion of the Universe as the end point of expansion, even
when \hbox{$\Lambda_\text{bare} = 0$}~\cite{Deffayet:2010qz,
Babichev:2012re, Bellini:2014fua}.

In this work we will be interested in classical nonlinear
regimes in these theories, where the higher-dimensional (HD)
operators $G_3(X)$ and/or $G_5(X)$ either balance or dominate
over the lower dimensional $G_2(X)$. This is the case in
cosmology and in the Vainshtein regime, respectively. From the
Effective Field Theory (EFT) point of view, the co-existence of
such a regime with the absence of strong coupling is possible, as
the nonlinearity energy scale is generically much smaller than
the strong-coupling scale, at which quantum effects become
important \cite{Heisenberg:2020cyi}. This is no different than
GR. Here, this is in part due to the nonrenormalization
properties enjoyed by Horndeski theories
\cite{Pirtskhalava:2015nla}. These two scales are background
dependent and their comparison should be done in a case by case
basis. Ultimately, as the strong-coupling scale is the naive
cutoff of the EFT (it may be lower), it should be kept
sufficiently large as to allow to describe the physics at the
scales of interest.

\subsection{Cosmological background}
\label{Sec2B}
Let us first consider an isotropic and homogeneous Universe
described by the Friedmann-Lema\^{\i}tre-Robertson-Walker (FLRW)
metric
\begin{equation} \label{FLRW}
ds^2 = -d\tau^2 +
a(\tau)^2\left(d\rho^2+\rho^2d\Omega^2\right).
\end{equation}
The field equations for the metric tensor give the value of
$H\equiv \dot a/a$ in terms of its sources.\footnote{Note that
this $H$ is defined in the ``Horndeski frame'', i.e., with
respect to the metric $g_{\mu\nu}$ of Eq.~(\ref{eqAction}). The
Jordan-frame (observable) Hubble expansion rate reads $e^{-\alpha
\varphi}(H+\alpha\dot\varphi)$, where $\alpha$ is the
matter-scalar coupling constant entering action~(\ref{eqAction}).
For $|\alpha|\lesssim 1$, such a change of frame does not modify
our order-of-magnitude estimates of Sec.~\ref{Observations}, and
the two frames strictly coincide when assuming $\alpha = 0$,
i.e., no direct matter-scalar coupling.} There may exist several
such sources, notably a bare cosmological constant
$\Lambda_\text{bare}$ as in action~(\ref{eqAction}), our Galileon
field $\varphi$, and possibly other fields. It will not be
necessary to write explicitly these equations for our study
below.

In the homogeneous Universe we consider, the scalar-field
equation of motion~\eqref{eq:phi-eom} reduces to
$\partial_\tau(a^3 J^0) = 0$, assuming no direct coupling to
other fields, therefore the shift-charge density $J^0 =
\text{const}/a^3$ tends towards $0$ during cosmological
expansion. The asymptotic cosmological solution for the time
derivative of the scalar field, say $\dot\varphi_c$, is thus
given by $J^0 = 0$ with
\begin{equation}
\frac{J^0}{M_\text{Pl}^2} = -2 k_2 \dot\varphi_c
+ \frac{6H}{M^2}\left[k_3 -
\left(\frac{H}{M}\right)^2k_5\right]\dot\varphi_c^2.
\label{J0}
\end{equation}
Aside from the trivial solution $\dot\varphi_c = 0$, for which
the scalar field does not contribute at all in the cosmological
expansion, and which is unstable owing to our choice of $k_2<0$,
there is an additional non-vanishing solution for $\dot\varphi_c$
when $k_2\neq 0$ and $k_3$ or $k_5$ do not vanish either. The
value of $\dot\varphi_c$ varies as an inverse power of $H$
as the Universe evolves, with $H$ given by the Friedmann
equation. The energy density carried by the homogeneous scalar
field in cosmology is in general given by
\begin{equation}
\rho_\varphi = -J^0\dot\varphi_c -2 k_2 \dot\varphi_c^2
+ \frac{2k_5}{M^4} H^3 \dot\varphi_c^3 \,.
\label{eq:rhophi}
\end{equation}
where the $J^0$ term disappears quickly as the shift charges
dilute. The value $\dot\varphi_c$ given by $J^0=0$ provides a
non-trivial contribution to the energy density, giving the
asymptotic value of $H$, which is then only a function of the
parameters $k_i$, $M$ and $\Lambda_\text{bare}$.

With $k_2 \sim -1$ and one of $k_{3,5} \sim 1$ and
$\Lambda_\text{bare} = 0$, we have $H \sim M$. We will refer to
this particular choice as \emph{full self-acceleration}, where
the scalar is alone responsible for driving the expansion of the
Universe.\footnote{In Secs.~\ref{Sec6} and \ref{Sec6C}, we shall
consider a model containing the three functions~(\ref{G235}) but
with $k_5$ small. The case $|k_2| \sim
|k_3| \sim M/H \sim 1$ will in any case correspond to full
self-acceleration.} This result exhibits the utility of using our
parametrization for the Horndeski functions~\eqref{G235}, for
which the dependence on $M_\text{Pl}$ factors out; in geometrical
units the equations would carry a small parameter. Retaining a
positive non-zero $\Lambda_\text{bare}\gg M^2$ implies that the
bare cosmological constant drives the acceleration. We stress
here the inversion that occurs with respect to the usual
situation: In our models a smaller $M$ implies a smaller effect
for the background, as opposed to the usual discussion where
large mass scales in operators suppress their effects. In
this context, the nonlinearity scale is $M$, while the naive
strong-coupling scale is $\Lambda_3 \sim (M_P M^2)^{1/3}$, i.e.,
the scale that suppresses the HD operators when expressed in
terms of canonically normalized fields, and clearly $M \ll
\Lambda_3$. Therefore this type of solution is well described
within the EFT.

We underline here the peculiarity resulting from the choice of
$k_2<0$, necessary for self-acceleration: there is a minimum
value of the scalar-field gradient. As it is approached, the
normalization of the acoustic (effective) metric for fluctuations
goes to zero, implying strong coupling.
Strong coupling is also present for a class of black hole
solutions in scalar-tensor
theories~\cite{Mukohyama:2005rw,Babichev:2018uiw,deRham:2019gha}.
It should be stressed that the solutions we consider below do not
belong to this ``stealth'' class, therefore we do not expect strong
coupling in our scenario.
In kinetic gravity
braiding, it was shown that there is a pressure singularity at
this point~\cite{Pujolas:2011he} on the other side of which the
scalar degree of freedom is a ghost~\cite{Babichev:2020tct}. A
similar mechanism should be present for generic Galileon theories
and is in fact a sign that the effective field theory description
is no longer valid. It is thus not possible to consistently
connect the time-like cosmological scalar-field gradient to a
static space-like gradient or even just a vacuum configuration
within the same theory.

We also note that the presence of $k_5$ modifies the speed of
propagation of gravitational waves on the cosmological
background~\cite{Bellini:2014fua}
\begin{equation}
\alpha_T \equiv c_T^2 - 1 =
\frac{2k_5\dot\varphi_c^3H}{M^4
- 2 k_5\dot\varphi_c^3 H}.
\label{eq:aT}
\end{equation}
This expression should also contain $\ddot\varphi_c$, which
vanishes for the future asymptotic background solution, and in
any case $\ddot\varphi \sim H\dot\varphi$ and we will neglect it.
The existence of such an operator is very strongly
constrained~\cite{Baker:2017hug,Creminelli:2017sry,
Ezquiaga:2017ekz,Sakstein:2017xjx}, by the measurement
of the speed of gravitational waves by
LIGO/Virgo~\cite{LIGOScientific:2017vwq}
\begin{equation}
\label{eq:aT0}
|\alpha_T^0| < 10^{-15},
\end{equation}
where we have added the superscript 0 to signify that this
constraint arises at low redshift and therefore for our models,
it implies that the deviation of GW speed from luminal must have
been even smaller in the past, as a result of the inverse
relationship between $\dot\varphi_c$ and $H$ implied by
Eq.~\eqref{J0}.

It is worth observing that, when the scale $M \sim H$ and the
scalar is fully responsible for the acceleration, the
corresponding strong-coupling scale $\Lambda_3$ is quite low and,
in terms of frequency, it lies just within the LIGO band. This
makes the EFT prediction of a nonluminal speed of gravitational
waves not robust for LIGO gravitational
waves~\cite{deRham:2018red}. It may well be that the speed of
tensors instead approaches the speed of light around the LIGO
band and therefore the above constraint would not be nearly as
strong, if it were at all.

\subsection{Test scalar-field generated by a black hole}
\label{Sec2C}
Let us now consider a static black hole of Schwarzschild radius
$r_S$ embedded within such an expanding Universe. Although the
scalar field may be responsible for this expansion, i.e., the
value of $H$, we assume that the \textit{local} scalar field
generated by the black hole is small enough not to
backreact\footnote{In a physically relevant scenario, the scalar
field either plays the role of the cosmological constant or is a
spectator field. Therefore, barring extra contributions from the
spatial derivatives ---which are suppressed due to the Vainshtein
mechanism when it \hbox{operates---,} one expects that the
backreaction is at most of order of that of the cosmological
constant, i.e., can be safely neglected at least locally. We
shall come back to this question below, notably when estimating
the scalar accretion rate in Sec.~\ref{sec:accretion-gen}.}
on the Schwarzschild-de~Sitter metric in static
coordinates\footnote{Related to the Friedmann coordinates of
Eq.~\eqref{FLRW} by the transformation \eqref{coord-change}.}
\begin{equation} \label{metric}
ds^2 = - f(r) dt^2 + \frac{dr^2}{f(r)} + r^2 d\Omega^2,
\end{equation}
where
\begin{equation}
f(r) = 1-\frac{r_S}{r} - (H r)^2.
\end{equation}
We look for a stationary solution of our test scalar field
in the form
\begin{equation}
\varphi = \dot\varphi_\text{BH}\, t +\phi(r),
\label{linearTime}
\end{equation}
where $\dot\varphi_\text{BH}$ is assumed to be
constant~\cite{Babichev:2012re,Babichev:2006vx,Babichev:2010kj,
Babichev:2013cya}. Since the radial derivative $\varphi' =
\phi'$, we shall actually no longer use the notation $\phi$ in
the following. The scalar-field equation $\nabla_\mu J^\mu = 0$
reads in such a case $\partial_r\left(r^2 J^r\right)= 0$,
therefore $r^2 J^r$ is given by an integration constant. In the
case of material bodies~\cite{Babichev:2012re}, writing this
equation within matter and integrating it over $r$ would imply
that this integration constant reads $\alpha r_S$ at lowest
post-Newtonian order, where $\alpha$ is the dimensionless
matter-scalar coupling constant entering the physical metric in
action~(\ref{eqAction}). In the present case of a black hole,
i.e., of a vacuum solution, the integration constant is not fixed
by any matter source, but rather by the regularity of the
solution. Let us denote it as $\alpha_\text{BH} r_S$ by analogy
with the matter case. The scalar-field equation reads then
\begin{equation}
\frac{J^r}{M_\text{Pl}^2} = \frac{\alpha_\text{BH} r_S}{r^2},
\label{EqJr}
\end{equation}
where
\begin{subequations}
\label{coefsABC}
\begin{eqnarray}
\frac{J^r}{M_\text{Pl}^2} &=& A \varphi'^2 + B \varphi' + C,
\label{Jr}\\
A &=& \frac{f}{M^2} \left[\left(\frac{4f}{r}
+ f' \right) k_3 +\frac{3 f - 1}{(M r)^2} f' k_5\right],
\label{coefA}\\
B &=& 2 f k_2,
\label{coefB}\\
C &=& -\left[ k_3 +\frac{f-1}{(M r)^2} k_5\right]
\frac{f' \dot\varphi_\text{BH}^2}{f M^2}.
\label{coefC}
\end{eqnarray}
\end{subequations}
This is thus a mere quadratic equation for $\varphi'$,
generalizing to $k_5 \neq 0$ the one derived for the cubic
Galileon case in~\cite{Babichev:2012re} (see
also~\cite{Babichev:2017guv,Lehebel:2018zga}).
Denoting its discriminant as
\begin{equation}
\Delta \equiv B^2
-4A\left(C-\frac{\alpha_\text{BH} r_S}{r^2}\right),
\label{discriminant}
\end{equation}
we have thus the very simple solution
\begin{equation}
\varphi' = \frac{-B\pm\sqrt{\Delta}}{2A}.
\label{scalarsolution}
\end{equation}
Note that this closed form is a consequence of our assumption of
linear Horndeski functions~(\ref{G235}). It will allow us to
analyze in detail its behavior at various locations close to the
black hole and at cosmologically large distances. Let us
immediately underline a crucial point: One must have $\Delta \geq
0$ for $\varphi'$ to be real. This means that when $\Delta$
reaches $0$ at a given radius $r$, it must have a \textit{double}
root to remain positive on both sides. We shall see below that
this actually provides the relationship between
$\alpha_\text{BH}$ and $\dot\varphi_\text{BH}$. In the full
domain, two such radii exist and this then actually fixes the
values of these two quantities. It is worth noting here that a
modification of $\dot\varphi_\text{BH}$ away from its
cosmological value $\dot\varphi_\text{c}$ does not prevent the
recovery of the homogeneous cosmological background at large
distances~\cite{Glavan:2021adm}
(see Appendix~\ref{sec:Homogeneity}).

Note also that the time derivative of the scalar field,
$\dot\varphi_\text{BH}$, enters in Eq.~(\ref{coefC}) as a second
source term for its radial derivative $\varphi'$, in addition to
the right-hand side of Eq.~(\ref{EqJr}). This consequence of the
nonlinearity of Horndeski theories implies that the cosmological
expansion has a direct effect on local solutions. This had
already been underlined for material bodies
in~\cite{Babichev:2012re,Anson:2020fum,Sakstein:2014isa,
Kobayashi:2014ida,Babichev:2016jom}, as well as for black holes
in other contexts~\cite{Babichev:2010kj}. As we will compute
later, the non-trivial background $\dot\varphi_\text{BH}$ induces
a scalar charge $\alpha_\text{BH}$ for the black holes with
$\alpha_\text{BH} \propto (\dot\varphi_\text{BH})^2$. Since in
our self-accelerating setup there exists a minimum value of
$\dot\varphi_\text{BH}$ for which the scalar-field fluctuations
are non-ghosts, this background cannot be removed completely and
a stationary black hole must always carry such a charge. We
expect a similar situation also for more general functions
$G_i(X)$, as a source term $C_0$ analogous to \eqref{coefC}, that
is a $\varphi'$-independent term, is always present for generic
$G_3(X)$ and $G_5(X)$ functions, namely \begin{eqnarray}
\label{general-source-term} C_0 = - \left[ G_{3}'(X) + \frac{f -
1}{r^2} G_{5}'(X) \right] \frac{f' \dot\varphi_\text{BH}^2}{f} .
\end{eqnarray} with both $G_{3}'(X)$ and $G_{5}'(X)$ evaluated at
$X = \dot\varphi_\text{BH}^2/f$. Interestingly, $G_2(X)$ and
$G_4(X)$ never contribute to it. We shall see below that it has
important observational consequences for black holes in the
present models.

\subsection{Accretion of scalar field}\label{sec:accretion-gen}

The ans\"atze we have assumed in Eqs.~\eqref{metric} and
\eqref{linearTime} may not be entirely consistent with one
another. While the shift-symmetry guarantees that the linear time
dependence of the scalar field does not appear in the equations
explicitly, the solutions to the combined field equations may
still show a nontrivial time evolution. This is due to a
nonvanishing energy flux through the black-hole horizon, as given
by the off-diagonal components of the stress-energy tensor,
\begin{equation}
\label{energy-flux}
T^r{}_t = - \dot\varphi_\text{BH} \, J^r |_{r=r_S},
\end{equation}
where we take the horizon to approximately be at $r = r_S$. This
form is general and a consequence of diffeomorphism
invariance~\cite{Babichev:2015rva}, and it implies that the
simultaneous presence of a time derivative of the scalar field
$\dot\varphi_\text{BH}$ and a non-zero shift-charge flux $J^r$,
gives rise to an energy flux. In Eq.~\eqref{EqJr} we see that the
shift-charge flux into the black hole is in turn proportional to
the scalar charge $\alpha_\text{BH}$. Note that there are
sub-classes of Horndeski theory which allow for solutions with a
time-dependent scalar~\eqref{linearTime} and zero energy flux,
see~\cite{Babichev:2013cya,Babichev:2016kdt}. For the solution we
consider here, however, there is always a non-zero accretion.

In order to trust our stationary ans\"atze, we must demand
that accretion of the scalar field into the black hole is a
sufficiently slow process. This condition may be expressed in
terms of the accretion rate associated with the above energy
flux (for details, see e.g.~\cite{Babichev:2012sg}),
\begin{eqnarray}
\label{acc-rate}
F_\text{acc}\sim r_S^2 \,
|T^r{}_t| = M_\text{Pl}^2 r_S |
\dot\varphi_\text{BH} \alpha_\text{BH} | ,
\end{eqnarray}
where we used Eq.~\eqref{EqJr} in the second equality.
Then using
\begin{equation}
\frac{d m}{dt} = \frac{M_\text{Pl}^2}2 \frac{dr_S}{dt}
= F_\text{acc},
\end{equation}
we find the characteristic time of the black hole mass change,
$\Gamma_\text{acc}^{-1}$,
\begin{equation}
\Gamma_\text{acc} \sim
|\dot\varphi_\text{BH}\, \alpha_\text{BH}|,
\label{Gammaacc}
\end{equation}
with $\alpha_\text{BH}\propto (\dot\varphi_\text{BH})^2$ in the
models we consider here, as we will demonstrate in the following
sections and as may already be anticipated from
equation~\eqref{coefC}.

Upon the formation of the black hole in the presence of the
cosmological background of the scalar field $\dot\varphi_c$, a
stationary solution is not possible without a charge appearing.
This is a result of the singularity in equation~\eqref{scalarsolution} which
appears unless we have a double root when $\Delta=0$ in the
vicinity of $r_S$. The black hole must then evolve on timescales
of order $\Gamma_\text{acc}^{-1}$ by accreting shift charge,
until this accretion is quenched, conservatively when the
configuration is such that
\begin{equation}\label{end-accretion}
\Gamma^{-1}_\text{acc} \sim H^{-1},
\end{equation}
the lifetime of the Universe.\footnote{Accretion can be
considered quenched when its characteristic time is longer than
the lifetime of the object. To be conservative, we assume that
this time is the lifetime of the Universe.} In the following, we
will estimate the potential for observability of such black holes
assuming this conservative state. In principle, the black holes
could be seen before they reach this final quasi-stationary state
and therefore with a larger charge and more hair.

The black hole then is in a quasi-stationary state. This happens
as a result of the two possible scenarios:
\begin{enumerate}
\renewcommand{\labelenumi}{\Roman{enumi}.}
\item \emph{Small accretion rate:} The cosmological
$\dot\varphi_c$ is such that the accretion rate is already slow
enough. The evolution of the black hole is effectively already
frozen and
\begin{equation}
\dot\varphi_\text{BH} \approx \dot\varphi_c.
\end{equation}
We will demonstrate that this is the scenario for the cubic
Galileon model.
\item \emph{Quenched accretion:}
Alternatively, if the charge induced by the cosmological
$\dot\varphi_c$ makes the accretion rate large, the black hole
instead begins to consume the energy stored in the scalar field's
cosmological background configuration, reducing the value of
$\dot\varphi$ in its vicinity. The accretion rate falls, but on
the timescale of the lifetime of the Universe, cannot decrease
below the inverse lifetime of the Universe. So, again
conservatively, the depletion would effectively freeze when
\begin{equation}
\dot\varphi_\text{BH} \ll \dot\varphi_c,
\end{equation}
reaching a value low enough so that Eq.~\eqref{end-accretion} is
satisfied. The charge of such a black hole would then be
disconnected from that implied by the cosmological background.
This is the scenario for sufficiently small black holes when a
quintic Horndeski operator is present.
\end{enumerate}

We stress that the alternative, maybe more usual, end point with
no scalar background or just a static spatial gradient is
\emph{not} a consistent solution that can be described by our
action. Scalar-field fluctuations on such backgrounds are ghosts
and the approach to the boundary between non-ghosts and ghosts
necessarily leads through a strong-coupling regime where
calculations are outside the validity of the effective field
theory describe large scales.

We should also remind the reader that $H$ is a function of time,
with $\dot\varphi_c$ given by Eq.~\eqref{J0} and therefore
smaller in the past. Our requirements on the accretion rate
should be interpreted as related to the time when the black hole
exists and e.g.~is emitting radiation. This means that higher
accretion rates in the past would be considered slow, while
charges would typically be smaller. We will show that the total
effect of this time-dependence is quite subtle and
model-dependent, as a result of a dependence of any observables
also on the screening.

\subsection{Vainshtein screening}
\label{VainshteinScreening}

As we will show in concrete models, black-hole charges induced by
cosmology in our class of theories are (very) large. This should
immediately appear as a potentially dangerous modification from
the standard situation, since it could spoil the experimental
tests of general relativity.

But the crucial difference with respect to the standard
scalar-tensor theories~\cite{Damour:1992we} (i.e., with a
canonical kinetic term (\ref{G2}) alone) is that there exists a
Vainshtein screening in the nonlinear Galileon theories, which
reduces the effect of the charge (see e.g.
Ref.~\cite{Babichev:2013usa} for a review on the Vainshtein
mechanism).

A good way to understand this is by noting that small-amplitude
high-frequency scalar perturbations effectively propagate in the
(inverse) acoustic metric $Z^{\mu\nu}$ and not the usual
spacetime (see e.g., Ref.~\cite{Sawicki:2024ryt} for a
pedagogical explanation of this phenomenon). When exchanging
scalar perturbations, the interaction between two bodies $A$ and
$B$ is proportional to the product of their scalar charges
$\alpha_A \alpha_B$, while the scalar propagator is built from
the inverse of the kinetic term. The interaction strength,
compared to that prevalent at cosmological scales away from
the body, is thus reduced by a multiplicative factor proportional
to
\begin{equation}
z_\lambda \equiv
\frac{Z_\lambda^{tt}}{Z_c^{tt}} \label{eq:zlambda}
\end{equation}
where $Z^{tt}_c \sim |k_2|$ is the acoustic metric normalization
at cosmological distances, $Z_\lambda^{tt}$ is the normalization
of the acoustic metric valid at the relevant scales, e.g.~the
wavelength of the gravitational radiation.
It is worth noting that the acoustic metric is not always
well defined, since the signature of the metric may be wrong and
in this case either gradient instability or a ghost may appear.
In particular, some black hole solutions with the configuration
of the scalar field $X = \text{const}$ indeed suffer from this
pathology~\cite{Khoury:2020aya,Takahashi:2021bml}. We stress,
however, that we consider here solutions that do not fall into
this class, since $X \neq \text{const}$. Moreover, in
Ref.~\cite{Babichev:2012re} it was demonstrated that close to the
material source, the perturbations in the $G_2+G_3$ model are
stable, which suggests that this may be also valid when a black
hole is considered in the same model.
A full stability analysis is postponed to future work.
\medskip

We will now present the method for estimating the relevant
$Z_\lambda^{tt}$. In the presence of derivative interactions
(i.e., non-canonical kinetic terms) the acoustic metric depends
on the background configuration of the scalar field. The
particular feature of the present Galileon models is that the
dependence on the scalar gradients is strong and the acoustic
metric changes significantly between that on the homogeneous
cosmological configuration and that in the vicinity of the black
hole. For the cubic Galileon, the metric is given in full
generality in Ref.~\cite[Eq.~(16)]{Babichev:2012re} (or
Ref.~\cite[Eq.~(2.15)]{Deffayet:2010qz}), where the relevant
terms involve $\varphi'$ and $\varphi''$.

A complication arises for Horndeski operators beyond $G_2(X)$,
namely the kinetic mixing between scalar and spin-2 degrees
of freedom. This is not problematic specifically for cubic
Horndeski, as for any $G_3(X)$ it is always possible to
decouple them generically for arbitrary
backgrounds~\cite{Babichev:2012re,Deffayet:2010qz,
Ezquiaga:2020dao}. However, for quintic Horndeski this is not
possible and the procedure instead becomes considerably more
difficult. Since we are only interested here in
order-of-magnitude estimates, in what follows we may proceed
without carrying out the diagonalization procedure, and instead
estimate separately the pure scalar part and the mixing term, as
a means of identifying the relevant overall scaling of the
effective metric for scalar perturbations.

In any case, the contributions involve the scalar field gradient
as sourced by the black hole. They will thus crucially depend on
the scalar charge $\alpha_\text{BH}$ induced by the cosmology,
giving a highly non-linear behavior.

For radii much larger than the Schwarzschild radius $r_S$
but much smaller than the cosmological horizon $1/H$,
the background scalar field solution around the black hole
(\ref{EqJr})-(\ref{coefsABC}) takes one of three forms, depending
on which term dominates the equation of motion. In $G_5$
dominance,
\begin{equation}
\varphi'^2 =
\frac{\alpha_\text{BH} M^4 r^2}{2 k_5}
\left[1 +\mathcal{O}\left(\frac{r_S}{r}\right)
+\mathcal{O}\left(H^2 r^2\right)\right],
\label{phiPrimeG5dominatedBis_gen}
\end{equation}
in regions where $G_3$ is most important,
\begin{equation}
\varphi'^2 =
\frac{\alpha_\text{BH} M^2 r_S}{4 k_3 r}
\left[1
+\mathcal{O}\left(\frac{r_S}{r}\right)
+\mathcal{O}\left(H^2 r^2\right)\right],
\label{phiPrimeG3dominated_gen}
\end{equation}
and for the $G_2$ kinetic term,
\begin{equation}
\varphi' =
\frac{\alpha_\text{BH} r_S}{2 k_2 r^2}
\left[1 +\mathcal{O}\left(\frac{r_S}{r}\right)
+\mathcal{O}\left(H^2 r^2\right)\right].
\label{phiPrimeG2dominatedBis_gen}
\end{equation}
As we will demonstrate in the subsequent analysis, the value of
$\alpha_\text{BH}$ depends on the particular model and its
parameters, but whatever it is, it creates the profiles
Eqs.~(\ref{phiPrimeG5dominatedBis_gen})--%
(\ref{phiPrimeG2dominatedBis_gen}).
Depending on the choices of the model parameters $k_i$ and $M$,
and the mass of the black hole, various terms will dominate in
different regions. Generically, $G_5$ dominates the solution at
small radii, $G_3$ at intermediate and $G_2$ at large radii.
However, parameters can be chosen where either $G_3$ or $G_5$ is
never the relevant solution (not in the least, when those
operators are absent from the model).

Comparing the order of magnitude of the
terms~(\ref{phiPrimeG5dominatedBis_gen})--%
(\ref{phiPrimeG2dominatedBis_gen})
allows us to define three different Vainshtein radii. The first,
typically the smallest one, occurs when
Eq.~(\ref{phiPrimeG5dominatedBis_gen}) is of the order
of~(\ref{phiPrimeG3dominated_gen}), and gives
\begin{equation}
r_{V35}^3 = \frac{|k_5| r_S}{2 |k_3| M^2},
\label{rV35}
\end{equation}
where the subscript ``$35$'' recalls that we are comparing the
$G_3$ and $G_5$-dominated expressions. An intermediate Vainshtein
radius can be defined when Eq.~(\ref{phiPrimeG5dominatedBis_gen})
is of the order of the square
of~(\ref{phiPrimeG2dominatedBis_gen}),
\begin{equation}
r_{V25}^3 = \frac{\sqrt{|k_5 \alpha_\text{BH}|}\,
r_S}{\sqrt{2} |k_2| M^2}.
\label{rV25bis}
\end{equation}
Finally, a third typically largest Vainshtein radius occurs when
Eq.~(\ref{phiPrimeG3dominated_gen}) is of the order of the square
of~(\ref{phiPrimeG2dominatedBis_gen}),
\begin{equation}
r_{V23}^3 =
\frac{|k_3 \alpha_\text{BH}| r_S}{k_2^2 M^2}.
\label{rV23}
\end{equation}
Note that the dependence of the Vainshtein radii on $M^2$ also
enters through $\alpha_\text{BH}$, which we will show is model
dependent.

\begin{figure}[t]
\includegraphics[width=0.48\textwidth]{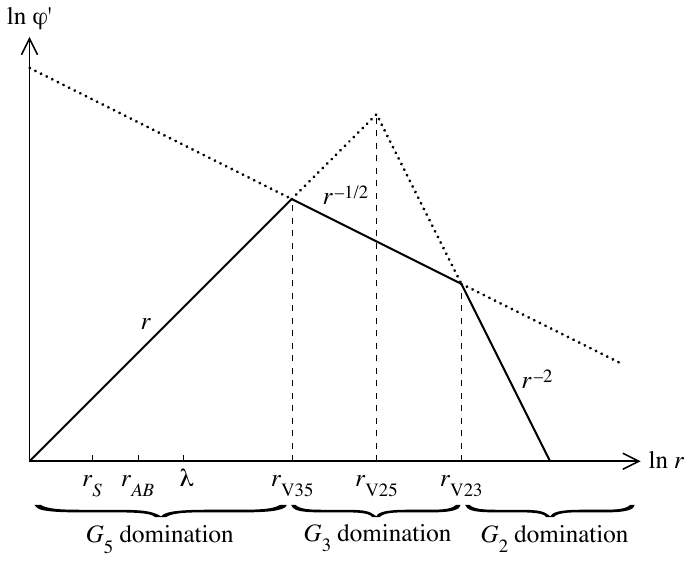}
\caption{The various regimes of the radial derivative $\varphi'$
in the quadratic $+$ cubic $+$ small quintic Galileon model for a
choice of black hole which is $G_5$ dominated. The gravitational
wavelength is denoted as $\lambda$, and $r_{AB}$ is an interbody
distance.}
\label{Fig1}
\end{figure}

The schematic behavior of the scalar field is illustrated in the
log-log plot of Fig.~\ref{Fig1}, with values calculated for an
example of model presented in section~\ref{Sec6}. Of course, the
actual solution has a smoother shape, and the cosmological
corrections $\mathcal{O}\left(H^2 r^2\right)$ we neglected start
to have an influence at large distances.

Given the solutions $\varphi'$ and the Vainshtein radii, we can
now calculate the acoustic metric suppression factors. The $G_2$
contribution to the metric is just $Z_2^{tt}\sim|k_2|$. The form
of the $G_3$ contribution to $Z^{\mu\nu}$ implies
that~\cite{Babichev:2012re}
\begin{equation}
\label{eq:Z3-gen}
Z_3^{tt}(r) \sim |k_2| \left(\frac{r_{V23}}{r} \right)^{3/2}.
\end{equation}
For the quintic Horndeski Lagrangian, at quadratic order in
scalar fluctuations, we find that the dominant contribution to the
purely scalar effective metric reads
\begin{eqnarray} \label{scalar-Z5}
Z_5^{\mu \nu} &\sim& \frac{k_5}{M^4} R^{\mu\alpha\nu\beta} \nabla_\alpha
\nabla_\beta \varphi, \quad
\end{eqnarray}
where $\varphi$ is the background scalar field. Evaluating the
$tt$ component, by inspection it is possible to see that there is
at least one derivative acting on the metric function $f$ (from
the curvature factor), providing a factor of $r_S$ in the
intermediate region far from $r_S$ and $H^{-1}$. Moreover the
whole expression itself is linear in $\varphi$, with at least one
derivative. The rest is dimensional analysis. Therefore, we can
estimate the leading contribution in this region to behave as
\begin{eqnarray}
Z_5^{tt} &\sim& \frac{k_5}{M^4} \frac{r_S}{r^4} \varphi'.
\label{Ztt}
\end{eqnarray}
Using the expression for $\varphi'$ in the nonlinear
regime~\eqref{phiPrimeG5dominatedBis_gen}, we find that
the acoustic metric behaves as
\begin{eqnarray}\label{Ztt52}
Z_5^{tt} &\sim& |k_2|
\left( \frac{r_{V25}}{r} \right)^3,
\end{eqnarray}
and therefore the Vainshtein screening is much stronger than for
$G_3$. The above estimate was obtained without accounting
for the kinetic mixing of the scalar with the gravitational
fluctuations. However, this is reliable as the mixing is
negligible in all the cases of interest that will be discussed in
the later sections (see Appendix~\ref{app:mixing}).

Whenever a particular $G_i$ lagrangian term dominates the
background, it also dominates the acoustic metric. We will thus
not need to distinguish in the remainder of this paper between
these two types of scales. We re-iterate here that $r_{V35}$
Eq.~\eqref{rV35} is independent of the black-hole charge
$\alpha_\text{BH}$, depending on the model parameters $k_i$, $M$
and the mass of the black hole only. A $G_5$-dominated region
surrounds a particular black hole only if $r_{V35}>r_S$.
Otherwise the black holes itself is $G_3$ dominated and it is the
$G_3$ term which can be seen to set the black hole charge.

If $G_5$ dominates for a particular black hole, the $G_3$ region
is transitory and does not affect the $Z_5$ term of the acoustic
metric beyond its intermediate effect on $\varphi'$. Note that
only $r_{V25}$ enters Eq.~\eqref{Ztt52}, and \emph{not}
$r_{V35}$, even though the location $r=r_{V25}$ does not
correspond to anything particular for the background solution
plotted in Fig.~\ref{Fig1}.

Even if the $G_5$ dominated region exists, a separate question is
the scale at which the acoustic metric is probed, $\lambda$. For
the emission of gravitational radiation in inspirals, this is the
wavelength of the GWs, i.e., $\lambda \sim 300\,r_S$ at the
moment the black-hole binaries enter the LIGO/Virgo band. This is
a somewhat larger scale than $r_S$ and therefore there exist
choices of parameters for which the black hole itself is in the
$G_5$ dominated region but where the radiation production is
determined by the $G_3$ term in the action. We will discuss this
in detail in section~\ref{Observations}.

Concerning the validity of the EFT to describe this very
nonlinear regime, it is interesting that the effect of the
Vainshtein screening in suppressing the interactions of the
scalar field, Eq.~\eqref{eq:zlambda}, tends to \emph{increase}
the strong-coupling scale well beyond the $\Lambda_3 \sim (M_P
M^2)^{1/3}$ that we have identified in the cosmological context.
On the other hand, we can identify a nonlinearity scale as
$r_V^{-1}$, for the largest of the Vainshtein radii. These makes
the two scales vastly different, making any such solutions well
described within the EFT.

\medskip

Having described the general features of the Galileon models, we
now turn to crux of this paper ---~the computation of the black
hole charges induced by cosmology, in the construction with
self-acceleration, $k_2<0$. We do this for two simpler subcases,
where the solutions and approximation are clear, and then
generalize to the full model involving both the cubic and quintic
Horndeski terms.

\section{Cubic Galileon: Small accretion}
\label{Sec4}

In the present section, we consider the particular case $k_5=0$
in Eqs.~(\ref{G235}), i.e., when only the two functions $G_2(X)$
and $G_3(X)$ define the dynamics of the scalar field.

The cosmological attractor Eq.~(\ref{J0}) in this model implies
\begin{equation}
\dot\varphi_c = \frac{k_2 M^2}{3 k_3 H},
\label{phiDotG2G3}
\end{equation}
which in turn gives the energy density of the scalar field
\begin{equation}
\label{eq:rhoG2G3}
\rho_\varphi =
-k_2 \dot\varphi_c^2 = \frac{|k_2|^3M^4}{9k_3^2 H^2}.
\end{equation}
This energy grows in time and therefore this kind of dark energy
is a phantom~\cite{Deffayet:2010qz}. Since $k_5=0$, the
propagation of gravitational waves is not modified in this model
and $\alpha_T=0$.

\subsection{The structure of the solution}
\label{Sec4A}

As is clear from Eq.~(\ref{scalarsolution}), there are two
possible branches of the solution, which correspond to the plus
and minus sign, respectively. The choice of the sign is
determined by the physical requirements on the scalar field
profile, i.e., one needs to choose the physically relevant
solution. To this end, let us first consider the scalar field
profile at large distances from a black hole. One expects to
recover the homogeneous scalar profile, i.e., a solution of the
scalar in a homogeneous FLRW Universe, with small corrections due
to the presence of the black hole. Setting $r_S=0$,
Eq.~(\ref{scalarsolution}) must therefore reduce to the
homogeneous solution, which in cosmological coordinates reads
\begin{equation}
\varphi = \dot\varphi_c \tau.
\label{phiBackgroundFriedmannCoords}
\end{equation}
The static and cosmological coordinates without the black hole
are related through
\begin{subequations} \label{coord-change}
\begin{eqnarray}
t &=& \tau - \frac{1}{2H} \log
\left[ 1 - \left( H e^{H\tau} \rho \right)^2 \right], \\
r &=& e^{H \tau} \rho \, ,
\end{eqnarray}
\end{subequations}
so that~\cite{Babichev:2012re}
\begin{equation}
\varphi' = -\dot\varphi_c\, \frac{Hr}{1-(Hr)^2}.
\label{phiPrimeBackground}
\end{equation}
Note that the homogeneous solution at large distances is
recovered from $\varphi'$, and \emph{not} from $\dot\varphi$ as
may have been naively expected. If we choose $\dot\varphi\neq
\dot\varphi_c$ in static coordinates, we would still recover the
cosmological solution, admittedly at the price of some
inhomogeneity decaying past the cosmological horizon (see
Appendix~\ref{sec:Homogeneity} for details).
\medskip

Still for $r_S=0$, the discriminant \eqref{discriminant} becomes a pure square,
$\Delta = [2 k_2 (1 - 3 H^2 r^2)/3]^2$, so that
solution~(\ref{scalarsolution}) reads
\begin{equation}
\varphi' =
-\dot\varphi_c H r\,
\frac{3 (1 - H^2 r^2)\pm
|1- 3 H^2 r^2|}{2 (1 - H^2 r^2)(2-3 H^2 r^2)}.
\end{equation}
To recover Eq.~(\ref{phiPrimeBackground}), one must choose $\pm =
\text{sign}(1- 3 H^2 r^2)$, i.e., plus for $H r < 1/\sqrt{3}$ and
minus for $H r > 1/\sqrt{3}$. The wrong sign would give
Eq.~(\ref{phiPrimeBackground}) divided by $(2-3 H^2 r^2)$,
corresponding to an inhomogeneous scalar background in FLRW
coordinates. The point $Hr=1/\sqrt{3}$, where the discriminant is
zero, is thus a branching point, where the sign entering
solution~(\ref{scalarsolution}) must change from $+$ to $-$ as
$r$ increases.

Moreover, the choice of the $-$ branch at very large radii, $H r
> 1/\sqrt{3}$, ensures that solution~(\ref{scalarsolution}) is
regular at the point where $A=0$ (i.e., $4f+r f'=0$ for the model
of the cubic Galileon, as one can read off Eq.~(\ref{coefA})),
which corresponds to $H r=\sqrt{2/3}$ for the homogeneous
solution. On the contrary, the choice of the $+$ branch would
lead to a singularity at $Hr=\sqrt{2/3}$. Indeed, close to this
radius, Eq.~(\ref{scalarsolution}) can be expanded as
\begin{equation}
\varphi' = \frac{|B|\pm|B|}{2 A}
\mp \frac{C}{|B|} + \mathcal{O}(A),
\end{equation}
showing that the upper sign is singular while the lower sign
gives a regular expression.

For a non-zero $r_S$, we expect a similar behavior of the
solution at large distances, with small corrections to the scalar
profile and the position of the cosmological branching point.

However, the presence of the black hole naturally introduces
another branching point in the vicinity of the black hole
horizon. Indeed, the $+$ branch of the solution, that we chose
above to match the cosmological background at smaller radii, $H r
< 1/\sqrt{3}$, diverges inside the Schwarzschild horizon at a
radius corresponding to $A=0$, namely at $r\simeq \frac{3}4r_S$
when neglecting the corrections due to the non-zero $H$. To avoid
this singularity at $A=0$ for small $r$, one needs to change
again to the $-$ branch at $\Delta=0$, similarly to our
description of the behavior of the solution at cosmologically
large distances.\footnote{One may think that such a singularity,
if present, may be disregarded, since it is inside the
Schwarzschild horizon. However, scalar perturbations in fact pass
through this horizon, as it generically happens for superluminal
perturbations~\cite{Babichev:2006vx}. Therefore a physically
relevant solution must be non-singular up to the horizon for
scalar perturbations, which lies inside the Schwarzschild
radius.} As we will confirm later, the branching point $\Delta=0$
is located at a slightly larger radius than the one for which
$A=0$, which allows the solution to avoid the singularity at
$A=0$.

The structure of the solution that we desire to construct is
summarized in Table~\ref{tab:table1}, while the details of
calculations will be given in the next subsections.

\begin{table}[t]
\caption{\label{tab:table1}
Special points of the test scalar-field solution in the quadratic
plus cubic Galileon model. We only display the first two terms of
the expansions in powers of the small quantity $H r_S$ (note that
the corrections are proportional to $H r_S^2$ for some terms but
to the much smaller $H^2 r_S^3$ for others). In the first line,
$Z^{\mu\nu}$ is the inverse effective metric in which scalar
perturbations propagate. The last column displays the sign to be
imposed in solution~(\ref{scalarsolution}) in order for it not to
deviate too much from the cosmological background $\varphi =
\dot\varphi_c \tau$.}
\begin{ruledtabular}
\begin{tabular}{llll}
Description & Equation & $r$ & $\pm \sqrt{\Delta}$\\
\colrule
sound horizon & $Z^{rr} = 0$
&$\frac{3}{4} r_S - \frac{9}{32} \sqrt{3} H r_S^2$
&$-$\\
pole & $4f+rf' = 0$ &
$\frac{3}{4} r_S + \frac{81}{128} H^2 r_S^3$
&$-$\\
branching point & $\Delta = 0$
&$\frac{3}{4} r_S + \frac{9}{128} \sqrt{3} H r_S^2$
&$\mp$\\
metric horizon & $f = 0$
&$r_S + H^2 r_S^3$
&$+$\\
\colrule
branching point & $\Delta = 0$
& $1/(\sqrt{3} H) + \frac{15}{2} r_S$
&$\pm$\\
pole & $4f+rf' = 0$ &
$\sqrt{2}/(\sqrt{3} H)-\frac{3}{8}r_S$
&$-$\\
metric horizon & $f = 0$
&$1/H - \frac{1}{2}r_S$
&$-$\\
\end{tabular}
\end{ruledtabular}
\end{table}

It is worth comparing the case of a black hole in cosmology with
the somewhat similar study of an asymptotically flat spacetime as
in Ref.~\cite{Babichev:2010kj}. Galileon accretion in
asymptotically flat spacetime requires the existence of a single
branching point, which is located in the vicinity of the BH
horizon. At this point, a branch of the solution which is
well-behaved at spatial infinity is matched to the branch
well-behaved near the horizon. In the case of accretion of a
perfect fluid, such a transition happens at the so-called
transsonic point. For a black hole in a FLRW Universe, as we have
seen above, there is an extra branching point at a cosmologically
large distance. This makes the construction of an everywhere
smooth and regular solution rather challenging, as each branching
point brings extra conditions on the profile of the solution. We
discuss this point in detail below.

\subsection{Branching points}
\label{Sec4B}

We underlined in Sec.~\ref{Sec2C} that the discriminant $\Delta$,
Eq.~(\ref{discriminant}), needs to remain positive or null at all
radii for our test scalar-field solution (\ref{scalarsolution})
to make sense. Since the locations of the roots of $\Delta$
depend on $\alpha_\text{BH}$ and $\dot\varphi_\text{BH}$, we must
define a procedure to fix these parameters so that these roots
are always \textit{double} roots. A single branching point
actually imposes a \textit{relation} between $\alpha_\text{BH}$
and $\dot\varphi_\text{BH}$, and one thus needs two branching
points to fix both of them. This allows us to construct a
quasi-stationary solution valid in the whole spacetime.
\medskip

Let us describe how $\alpha_\text{BH}$ can be determined in terms
of $\dot\varphi_\text{BH}$ by enforcing a double root at one
specific radius. To do so, we write $\Delta = N(r)/D(r)$ as a
ratio of polynomials depending on~$r$. Its numerator $N(r)$ is of
9th degree. Without yet knowing the radius $r = r_\text{root}$
where $N(r)$ vanishes, we impose that its radial derivative must
also vanish at the same location. We have thus a set of two
polynomial equations, $N(r_\text{root}) = 0$ and
$N'(r_\text{root}) = 0$. One of them may be used to write
$\alpha_\text{BH}$ in terms of the (still unknown)
$r_\text{root}$, and be replaced in the other equation. This
gives now a polynomial of 14th degree. Since its exact real roots
cannot be written in a closed form, one then looks for them in a
perturbative way, by increasing progressively the relative power
of the very small quantity $H r_S$ up to which the solution is
correct.\footnote{$H r_S \sim 10^{-22}$ for the typical black
holes observed in the LIGO/Virgo interferometers, and $\sim
10^{-18}$ to $10^{-16}$ for the heavy ones expected to be
detected with the LISA mission.} More specifically, we start by a
trial value $r_\text{trial}$ and look for a next approximation in
the form $r_\text{trial} +\delta r$, that we insert within the
14th-degree polynomial which must vanish. Solving for this small
$\delta r$ (at first order, or at second order for the first
step) gives the next trial value, and we iterate this procedure.
Once $r_\text{root}$ has been determined with enough precision,
say up to relative order $\mathcal{O}\left(H^n r_S^n\right)$, one
can check that both $N(r_\text{root})$ and $N'(r_\text{root})$
vanish, but that the second derivative $N''(r_\text{root})$ does
\textit{not} (otherwise this would be a third root and the
discriminant $\Delta$ could become negative). More specifically,
if $N(r)\propto (r - r_\text{root})^2$, then $N'(r)\propto (r -
r_\text{root})$, therefore $N(r_\text{root})$ should be of order
$\mathcal{O}\left(H^{2n} r_S^{2n}\right)$, $N'(r_\text{root})$ of
order $\mathcal{O}\left(H^n r_S^n\right)$, while
$N''(r_\text{root})$ should not have any such factor. Finally,
when $r_\text{root}$ has been obtained from this procedure, one
can replace it in the expression of $\alpha_\text{BH}$, and this
fixes in a unique way the integration constant $\alpha_\text{BH}
r_S$ entering Eq.~(\ref{EqJr}), in terms of
$\dot\varphi_\text{BH}$.

Repeating the same procedure at the second branching point
generates another relation, and their combination allows us to
determine both $\alpha_\text{BH}$ and $\dot\varphi_\text{BH}$.

\subsection{Branching point close to the black hole}
\label{Sec4C}
When looking for such a double root of $\Delta$ at small radii of
the order of the Schwarzschild radius $r_S$, and assuming that
$\dot\varphi_\text{BH}$ is of the order of magnitude of the
cosmological background value \eqref{phiDotG2G3}, one finds
\begin{subequations}
\label{rootalpha1}
\begin{eqnarray}
r_\text{root} &=& \frac{3}{4} r_S
\left[1+ \frac{\sqrt{3} k_2 M^2}{32 k_3
\dot\varphi_\text{BH}} r_S
+ \mathcal{O}\left(H^2 r_S^2 \right)\right],
\label{root1}\\
\alpha_\text{BH} &=& 3 k_3
\left(\frac{\dot\varphi_\text{BH}}{M}\right)^2
+\frac{3\sqrt{3}}{4} k_2\dot\varphi_\text{BH} r_S
+ \mathcal{O}\left(H^2 r_S^2\right).\nonumber\\
\label{alphaBH1}
\end{eqnarray}
\end{subequations}
To simplify, we only write the first two terms in the expansions,
but we did compute several orders more to check the behaviors of
$N(r_\text{root})$, $N'(r_\text{root})$ and $N''(r_\text{root})$
as explained above.

With the assumption that $\dot\varphi_\text{BH} = \dot\varphi_c$
strictly, this gives
\begin{subequations}
\label{rootalphac}
\begin{eqnarray}
r_\text{root}^{(c)} &=& \frac{3}{4} r_S
\left[1+ \frac{3 \sqrt{3}}{32} H r_S
+ \mathcal{O}\left(H^2 r_S^2 \right)\right],
\label{rootc}\\
\alpha_\text{BH}^{(c)} &=&
\frac{1}{3 k_3}\left(\frac{k_2 M}{H}\right)^2
\left[1
+ \frac{3\sqrt{3}}{4} H r_S
+ \mathcal{O}\left(H^2 r_S^2\right)
\right],
\nonumber\\
\label{alphaBHc}
\end{eqnarray}
\end{subequations}
where the upper index $(c)$ recalls this assumption. When the
Galileon field $\varphi$ is responsible alone for cosmological
expansion $\alpha^{(c)}_\text{BH}$ is maximal. In that case,
Ref.~\cite{Babichev:2012re} showed that $H^2 = (|k_2|/3)^{3/2}
M^2/|k_3|$, and we get
\begin{equation}
\frac{\alpha_\text{BH}^{(c)}}{\text{sign}(k_3)\sqrt{-k_2}} =
\sqrt{3} + \frac{9}{4} H r_S
+ \mathcal{O}\left(H^2 r_S^2\right).
\label{alphaBHc2}
\end{equation}
Let us recall that it is always possible to choose $k_2 = -1$
(or $-\frac{1}{2}$, depending on the reader's preferences)
by reabsorbing it in the definition of $\varphi$ in action
(\ref{eqAction}). The important point to already note here is
that this dimensionless scalar charge is of order $1$,
independently of any direct matter-scalar coupling constant
$\alpha$ which may have been assumed in action (\ref{eqAction}),
and even if $\alpha = 0$ strictly. In other words, even if one
assumes that the scalar field is not directly coupled at all to
any matter, the regularity of the solution implies that black
holes must be significantly coupled to it. This surprising effect
is a result of needing to avoid a potential singularity in the
solution close to the black hole $r=r^{(c)}_\text{root}$,
Eq.~\eqref{rootc}, in the presence of $\dot\varphi_\text{BH}$.
Despite being driven by the cosmological background, it is a
local effect which forces the relation~(\ref{alphaBH1}).

In asymptotic Minkowski spacetime with $\dot\varphi_c = 0$, pure
Schwarzschild black holes without any scalar hair ($\varphi
=\text{const}$ everywhere) are solutions of the field equations.
Here, this is because Eq.~(\ref{coefC}) behaves as a source for
$\varphi'$ that we find such a hair.

\subsection{Small accretion rate}
Let us check that this configuration is the (approximate) end
point of the evolution of the black hole, as we discussed in
section~\ref{sec:accretion-gen}. The accretion rate for the
charge~\eqref{alphaBHc} can be estimated as
\begin{eqnarray}
\Gamma_\text{acc} \approx \frac{|k_2|^3 M^4}{9 |k_3|^2 H^3}
= \frac{\rho_\varphi}{H} \lesssim H ,
\end{eqnarray}
where we have also used Eq.~\eqref{eq:rhoG2G3} for the energy
density of the scalar $\rho_\varphi$. Thus for this solution,
scalar-field accretion into the black hole is negligible and the
charges are long lived. We are within scenario I of
section~\ref{sec:accretion-gen}. Note that the accretion rate in
this model is even smaller in the past, $\rho_\varphi \propto
H^{-2}$.

We may also estimate the contribution of the energy-momentum
tensor of the scalar field close to the black hole, to check that
our assumed background Schwarzshild-de~Sitter
metric~(\ref{metric}) is not significantly affected, and
therefore that our computation of the scalar charge
$\alpha_\text{BH}$ can be trusted. It is dominated by the $G_3$
term entering action~(\ref{eqAction}), of order $k_3 X \Box
\varphi/M^2$. Its integral within a sphere of radius a few times
$r_S$ gives a result of order $\mathcal{O}(H r_S^2)$, therefore
extremely small with respect to the assumed $r_S$ in
metric~(\ref{metric}). This is another way to confirm that the
scalar's backreaction is negligible.

\subsection{Branching point at large distance}
\label{Sec4D}

As we have already demonstrated in section~\ref{Sec4A}, there
exists a second branching point at a cosmologically large
distance $r\approx 1/(\sqrt{3}H)$ from the black hole. We show in
Appendix~\ref{sec:G3-cosmo-branch} that when assuming
$\dot\varphi_\text{BH} = \dot\varphi_c$ strictly, it is strongly
inconsistent with the local branching point studied in
Sec.~\ref{Sec4C} above. It would indeed need a small
$\mathcal{O}(H r_S)$ scalar charge, instead of the large
$\mathcal{O}(1)$ value derived in Eqs.~(\ref{alphaBHc}) or
(\ref{alphaBHc2}). In order to make these two branching points
consistent with each other, we must now tune our second free
parameter, $\dot\varphi_\text{BH}$, which may slightly differ
{}from $\dot\varphi_c$.

This discussion is actually valid for any scalar charge, even the
bare one $\alpha$ assumed for matter in action (\ref{eqAction}).
Let us thus drop for a while the subscript ``BH'', and assume
\begin{equation}
\dot\varphi_\text{local} = \dot\varphi_c
\times
\left[1 + \kappa H r_S
+\mathcal{O}\left(H^2 r_S^2\right)\right],
\end{equation}
where $\kappa$ is an $\mathcal{O}(1)$ dimensionless parameter
we wish to determine. Using again the perturbative technique
described in Sec.~\ref{Sec4B}, we now assume that $\alpha$
is fixed (either by the action for matter bodies, or from the
regularity of the local solution for black holes), and we look
for $\dot\varphi_\text{local}$ such that the discriminant
$\Delta$ admits a double root near $r\approx 1/(\sqrt{3}H)$.
One finds
\begin{subequations}
\label{phiDotRootConsistent}
\begin{eqnarray}
\dot\varphi_\text{local} &=& \dot\varphi_c
\times
\left[1 +
\frac{9 \sqrt{3}\, k_3 H^2}{2 k_2^2 M^2} \alpha H r_S
+\mathcal{O}\left(H^2 r_S^2\right)\right],
\nonumber\\
\label{phiDotLocal}\\
r_\text{root} &=&
\frac{1}{\sqrt{3}\, H} + \left(\frac{3}{2}
+ \frac{18 k_3 \alpha H^2}{k_2^2 M^2}\right)r_S
+ \mathcal{O}\left(H r_S^2 \right).
\nonumber\\
\label{rootConsistent}
\end{eqnarray}
\end{subequations}
Note that the corrections to $\dot\varphi_c$ and
$1/(\sqrt{3}H)$ are extremely small, even for a scalar
charge $\alpha$ of order 1, since $H r_S \sim 10^{-22}$.
In conclusion, our assumption of a linear time-dependent test
scalar field of the form~(\ref{linearTime}) is now consistent at
all radii, even for the case of matter bodies considered in
Ref.~\cite{Babichev:2012re}. The only price to pay is a tiny
modification of the local time derivative of the scalar field
with respect to its asymptotic cosmological value. In the case of
black holes we are considering in the present paper, the
conclusion is that such a tiny modification of
$\dot\varphi_\text{BH}$ suffices to make both branching points
near $r\approx \frac{3}{4} r_S$ and $r\approx 1/(\sqrt{3}H)$
consistent with each other. Of course, this change of
$\dot\varphi_\text{BH}$ also implies very small modifications
with respect to Eqs.~(\ref{rootalphac}). Our final results read
\begin{subequations}
\label{correct23}
\begin{eqnarray}
\dot\varphi_\text{BH} &=& \dot\varphi_c \times
\left[1 + \frac{3}{2}\sqrt{3} H r_S
+\mathcal{O}\left(H^2 r_S^2\right)\right],
\label{correctPhiDot}\\
r_\text{root}^\text{close} &=&
\frac{3}{4} r_S
\left[1+ \frac{3 \sqrt{3}}{32} H r_S
+ \mathcal{O}\left(H^2 r_S^2 \right)\right],
\label{correctRoot1}\\
r_\text{root}^\text{far} &=&
\frac{1}{\sqrt{3}\, H} + \frac{15 r_S}{2}
+ \mathcal{O}\left(H r_S^2 \right),
\label{correctRoot2}\\
\alpha_\text{BH} &=& \frac{1}{3 k_3}
\left(\frac{k_2 M}{H}\right)^2
\left[1 + \frac{15}{4} \sqrt{3} H r_S
+\mathcal{O}\left(H^2 r_S^2\right)\right].
\nonumber\\
\label{correctAlpha}
\end{eqnarray}
\end{subequations}
Here again, we have actually computed several orders more in
these expansions, but we only display the first two terms in each
equation. Note that at this order, the position of the close
root, Eq.~(\ref{correctRoot1}), did not change with respect to
(\ref{rootc}). On the other hand, the second terms of
Eqs.~(\ref{correctAlpha}) and (\ref{correctRoot2}) are five times
larger than what we obtain while assuming $\dot\varphi_\text{BH}
= \dot\varphi_c$ strictly in Eq.~(\ref{alphaBHc}) above and in
Eq.~(\ref{rootinconsistent}) of
Appendix~\ref{sec:G3-cosmo-branch}. Let us also quote the value
of the scalar charge when the Galileon field is responsible alone
for the cosmological expansion, i.e., the corrected version of
Eq.~(\ref{alphaBHc2}):
\begin{equation}
\frac{\alpha_\text{BH}}{\text{sign}(k_3)\sqrt{-k_2}} =
\sqrt{3} + \frac{45}{4} H r_S
+ \mathcal{O}\left(H^2 r_S^2\right).
\label{correctAlpha2}
\end{equation}

We should underline that the above modification of
$\dot\varphi_\text{BH}$ with respect to $\dot\varphi_c$ is
consistent with our assumption (\ref{linearTime}) of a linear
time dependence of the scalar field everywhere in spacetime, but
that other solutions are also possible. From a physical
viewpoint, the local fields must react quickly to avoid any
singularity, but the inconsistency we find in
Appendix~\ref{sec:G3-cosmo-branch} at the cosmologically large
distance $r\approx 1/(\sqrt{3}H)$ needs much more time to
backreact on $\dot\varphi_\text{BH}$. We can thus argue that in
the realistic setup of the formation of a black hole, for
instance from the collapse of matter not directly coupled to the
Galileon (i.e., $\alpha = 0$ action~(\ref{eqAction})), the actual
scalar field should have a more complex time dependence
than~(\ref{linearTime}). Once it is formed in a background with a
time derivative of the scalar field equal to $\dot\varphi_c$, it
quickly stabilizes to avoid local singularities, therefore it
adjusts its scalar charge to Eq.~(\ref{alphaBHc}) so that the
discriminant $\Delta$ has a double root near $r\approx
\frac{3}{4} r_S$. Its scalar hair then propagates at a finite
velocity towards large radii, and when it reaches $r\approx
1/(\sqrt{3}H)$, the field equation ``realizes'' that
$\dot\varphi_\text{BH}$ needs to be adjusted for a double root of
$\Delta$ to also exist there. It then sends back this
information, again at a finite velocity, towards the location of
the black hole. Therefore, one may argue that the imprecise
scalar charge (\ref{alphaBHc}) has probably more physical meaning
than the correct one (\ref{correctAlpha}) needed for an exact
linear time dependence everywhere. Moreover, as soon as there
exist several black holes of different masses in the Universe, as
well as the possibly coupled matter bodies ($\alpha\neq 0$ in
action~(\ref{eqAction})), then a uniform $\dot\varphi$ is
obviously no longer possible everywhere. But since all local
$\dot\varphi_\text{BH}$ and the background $\dot\varphi_c$ only
differ by a tiny amount of relative order $\mathcal{O}(H r_S)\sim
10^{-22}$, this does not change anything to the observational
consequences discussed in Sec.~\ref{Observations}.

\subsection{Vainshtein screening}
\label{Sec4E}
For the cubic Galileon model, the relevant Vainshtein radius is
given by Eqs.~(\ref{rV23}) and (\ref{correctAlpha}):
\begin{equation}
r_{V23}^3 =
\frac{k_3 \alpha_\text{BH} r_S}{k_2^2 M^2}
\approx \frac{r_S}{3 H^2}.
\label{rV23cubic}
\end{equation}
It is interesting to note that it only depends on the physical
quantities $r_S$ and $H$, but no longer on any parameter entering
action~(\ref{eqAction}), i.e., $k_2$, $k_3$ nor $M$. In
particular, it keeps strictly the same value even if the Galileon
field is not responsible alone for the accelerated expansion of
the Universe, and even if $M\ll H$. This independence of the
Vainshtein radius from the theory parameters comes from the fact
that we are considering only two linear Galileon kinetic
terms~(\ref{G235}). It generically does depend on $M$ in other
models, as Sec.~\ref{Sec5C} will illustrate.

Substituting Eq.~\eqref{rV23cubic} into Eq.~\eqref{eq:Z3-gen},
we obtain
\begin{equation}\label{Z3-model1}
Z_3^{tt} \sim \frac{|k_2|}{\sqrt{3}Hr_S}
\left(\frac{r_S}{r}\right)^{3/2}.
\end{equation}
We discuss the observational consequences of the combination of
the large scalar charge~\eqref{correctAlpha} and the acoustic
metric~\eqref{Z3-model1} in Sec.~\ref{Sec6Aobs}.

\section{Simplest quintic Horndeski term}
\label{Sec5}

\subsection{Scalar field solution}
\label{Sec5A}

Let us now consider the particular case $k_3=0$ in
Eqs.~(\ref{G235}), i.e., when only the two functions $G_2(X)$ and
$G_5(X)$ define the dynamics of the scalar field. In such a case,
Eq.~(\ref{J0}) imposes
\begin{equation}
\dot\varphi_c = -\frac{k_2 M^4}{3 k_5 H^3},
\label{phiDotG2G5}
\end{equation}
instead of Eq.~(\ref{phiDotG2G3}).
This theory is very simple and the energy density of the scalar
\eqref{eq:rhophi} can be directly related to the speed of
gravitational waves~\eqref{eq:aT},
\begin{align}
\frac{\rho_\varphi}{12H^2} = \alpha_T =
\frac{2}{27} \frac{|k_2|^3}{k_5^2}
\left(\frac{M}{H}\right)^8.
\label{eq:M2_aT}
\end{align}
The constraint~\eqref{eq:aT0} then implies that such a model
cannot drive the acceleration of the Universe. If we choose
$|k_2| \sim |k_5| \sim 1$ which can be reabsorbed in the
definitions of $\varphi$ and $M$, cf.~Eqs.~(\ref{G235}), this
translates as a limit on $M/H \lesssim 2\times 10^{-2}$, still a
mild hierarchy of scales. Note also the $H^{-8}$ dependence in
expression~\eqref{eq:M2_aT}: However small $\alpha_T$ is allowed
to be today, in this model it would have been much smaller in the
recent past.

We follow the same procedure as described in Sec.~\ref{Sec4B} to
determine the precise location of a double root of the
discriminant $\Delta$ close to the black hole, and the
corresponding scalar charge $\alpha_\text{BH}$. A difference is
that when writing the discriminant $\Delta = N(r)/D(r)$ as a
ratio of polynomials, its numerator $N(r)$ is now of 14th degree.
And when replacing the expression of $\alpha_\text{BH}$ imposed
by one of the two equations $N(r_\text{root}) = 0$ or
$N'(r_\text{root}) = 0$ into the other, we now get a polynomial
of 23rd degree. The perturbative search for such a double root
is however similar, and we now find that it must be close to
$r_\text{root} \approx \frac{3}{2} r_S$ (as compared to
$\frac{3}{4} r_S$ in Sec.~\ref{Sec4C}).

We then find that the analogues of Eqs.~(\ref{rootalphac}) read
in the present quadratic plus quintic model
\begin{subequations}
\label{rootalphac25}
\begin{eqnarray}
r_\text{root}^{(c)} &=& \frac{3}{2} r_S
\left[1+ \frac{27}{8} H^2 r_S^2
+ \mathcal{O}\left(H^3 r_S^3 \right)\right],
\label{rootc25}\\
\alpha_\text{BH}^{(c)} &=&
\frac{2}{k_5}\left(\frac{2 k_2 M^2}{9 H^3 r_S}\right)^2
\left[1
- \frac{27}{2} H^2 r_S^2
+ \mathcal{O}\left(H^3 r_S^3\right)
\right],
\nonumber\\
\label{alphaBHc25}
\end{eqnarray}
\end{subequations}
where the upper index $(c)$ recalls our assumption
$\dot\varphi_\text{BH} = \dot\varphi_c$. The crucial difference
with Sec.~\ref{Sec4} is that the scalar charge is now
proportional to $1/(H r_S)^2$, as compared to $\mathcal{O}(1)$ in
Eqs.~(\ref{alphaBHc}) or (\ref{alphaBHc2}). In other words, in
this model, the requirement of stationarity of the solution in
the presence of a non-vanishing $\dot\varphi_c$ implies extremely
large scalar charges for black holes. This comes from the
coefficient $1/(M r)^2$ entering Eq.~(\ref{coefC}). Moreover, the
dimensionless charges now depend on the Schwarzschild radius
$r_S$, therefore black holes of different masses have different
scalar charges, so that dipolar radiation becomes possible. As an
illustration, taking $H r_S \sim 10^{-22}$ (for LIGO black
holes), we could thus expect scalar charges as high as order
$10^{42}$ in the fully self-accelerated case $|k_2|\sim |k_5|
\sim M/H \sim 1$.

Attempting to correct this solution by taking account of the root
at cosmological distances as in the cubic Galileon case does not
produce a good solution. A second branching point near $r\approx
1/(\sqrt{3}H)$ should exist. We can thus let the local
$\dot\varphi_\text{BH}$ differ from the background
$\dot\varphi_c$, and look as before for the value which would
allow for a double root of the discriminant $\Delta$ near this
radius. We find that this is actually impossible, because the
existence of such a double root would need a \textit{negative}
value of the square $\dot\varphi_\text{BH}^2$. Our
assumption~(\ref{linearTime}) of a linear time dependence
everywhere is inconsistent. As we will demonstrate, the accretion
rate implied by the cosmological background in this model is
typically large and does not support the type of stationary
solution described in Sec.~\ref{Sec4D} valid everywhere in
spacetime (the solution would have evolved on the timescale
required to adjust to the root at cosmological distances). As
argued previously, fixing the behavior at small distances is
sufficient for our discussion of the local physics of radiation.

\subsection{Accretion scenarios}
\label{Sec5LargeAccretion}

Given the large scalar charge for black holes in this model, one
may expect that the accretion rate can be rather large. Let us
make a connection with possible observations by assuming that we
have a given fixed black hole mass represented by its $r_S$. For
the accretion rate~\eqref{Gammaacc} onto such a black hole to be
slow, $\Gamma_\text{acc}^{(c)}\lesssim H$, the expressions for
$\dot\varphi_c$ Eq.~\eqref{phiDotG2G5} and $\alpha_\text{BH}$
Eq.~\eqref{alphaBHc25} imply that we require
\begin{equation}
\frac{M}{H} \lesssim
\left(\frac{3^5 k_5^2}{2^3 |k_2|^3} H^2 r_S^2\right)^{1/8},
\label{smallMH}
\end{equation}
In other words,
\begin{equation}
\Gamma^{(c)}_\text{acc} =
\frac{2^3 |k_2|^3 M^8}{3^5 k_5^2 H^9 r_S^2}
\approx \left(\frac{2}{3 H r_S}\right)^2
\alpha_T H,
\label{g25-slow-acc}
\end{equation}
where we have used Eq.~\eqref{eq:M2_aT}, and $\alpha_T$ scales as
$H^{-8}$. The requirement for small accretion in this model then
implies that $\alpha_T<(3Hr_S/2)^2\sim 10^{-43}$ for LIGO/Virgo
black holes with masses $\sim 10\, m_\odot$, and
$\alpha_T<10^{-35}$ for LISA supermassive black holes with masses
$\sim10^5 m_\odot$. For $M/H$ larger than implied by these
limits, black holes of sizes that we will observe would not be in
the small accretion scenario. However, since
$\Gamma_\text{acc}^{(c)}/H \propto H^{-10}$, the range of
black-hole masses which are accreting slowly on the cosmological
background is larger at high redshift in this model ---~for a
source at redshift $z=2$ these above conditions are relaxed by a
factor of $[H(z=2)/H(z=0)]^{10}\sim 10^{5}$, and $\sim10^{9}$ for
a source at redshift $z=5$.

Indeed, for any value of the parameters of the action,
sufficiently small black holes, $(H r_S)^2<\alpha_T$, are in the
quenched accretion scenario of~\ref{sec:accretion-gen}. When such
black holes form on the cosmological background $\dot\varphi_c$,
the accretion rate is initially large and they absorb the
scalar-field background reducing the local $\dot\varphi$ and
therefore their charge. The asymptotic future configuration is
decoupled from the cosmological background. Once the timescale of
accretion reduces to $\Gamma_\text{acc} \ll r_S^{-1}$, we can
think of the quasi-stationary configuration, determined by
requiring the presence of the double root at
$r\approx\frac{3}{2}r_S$ given some $\dot\varphi_\text{local}$,
as being the approximate solution which evolves adiabatically.
In such a case we have
\begin{equation}
\alpha_\text{BH} \approx 2k_5\left(\frac{2\,
\dot\varphi_\text{local}}{3M^2r_S}\right)^2.
\label{alphaLocal}
\end{equation}
The accretion rate will continue to fall as the scalar background
is absorbed, but it cannot go below that implied by the lifetime
of the Universe, $\Gamma_\text{acc}\gtrsim H$. We thus have a
\emph{lower} bound for the local value of the scalar derivative
and the charge in this model, both dependent on $r_S$,
\begin{subequations}
\begin{align}
|\dot\varphi_\text{local}| &\gtrsim
\left(\frac{9 H M^4 r_S^2}{8 |k_5|}\right)^{1/3},
\label{phi25-low}\\
|\alpha_\text{BH}| &\gtrsim
2\left(\frac{|k_5| H^2}{9 M^4 r_S^2}\right)^{1/3}.
\label{alpha25-low}
\end{align}
\end{subequations}

Depending at which point in their evolution these black holes are
observed, the charge will be somewhere between the initial value
implied by the cosmological background~\eqref{alphaBHc25} and the
lower bound~\eqref{alpha25-low}. This is a generic feature of
such a quintic operator, and in the subsequent we will assume
that the charge is at this lower bound for all black holes which
have undergone the quenched accretion scenario. This is a very
conservative estimate.

When they are in this quenched state, one may also estimate
the contribution of the energy-momentum tensor of the scalar
field close to the black hole. It is dominated by the $G_5$ term
entering action~(\ref{eqAction}), of order $k_5 X \varphi''
f''/M^4$. Its integral within a sphere of radius a few times
$r_S$ gives again a result of order $\mathcal{O}(H r_S^2)$,
therefore negligible with respect to the assumed $r_S$ in
metric~(\ref{metric}).

\subsection{Vainshtein screening}
\label{Sec5B}
As we have already previewed in Sec.~\ref{VainshteinScreening},
in the intermediate range $r_S \ll r \ll 1/H$, the scalar field
solution (\ref{EqJr})-(\ref{coefsABC}) is dominated by its $G_5$
kinetic term at small distances and the $G_2$ term at large
distances, giving the approximate solutions
\eqref{phiPrimeG5dominatedBis_gen} and
\eqref{phiPrimeG2dominatedBis_gen} respectively. Assuming that
our black hole was formed large enough to accrete slowly, the
Vainshtein radius determined by its charge is universal for all
such black holes and given by Eq.~\eqref{rV25bis}
\begin{equation}
r_{V25}^3 = \frac{\sqrt{k_5 \alpha_\text{BH}}\,
r_S}{\sqrt{2}\, |k_2| M^2}
= \frac{2}{9 H^3},
\label{rV25-model2}
\end{equation}
where $\alpha_\text{BH}$ is given by Eq.~(\ref{alphaBHc25}),
i.e., a very large radius of a size comparable to the observable
Universe.\footnote{Note that in the present model again, this
Vainshtein radius does not depend on the theory parameters, but
only on the physical quantity $H$ (and not even $r_S$, here).}
Obviously, our assumption $r \ll 1/H$ is not satisfied at such a
large radius, but this anyway shows that coalescences of such
black holes happen deep within their Vainshtein region.

On the other hand, when the black hole initially accretes
quickly, the quenched charge at the time of observation is much
reduced. This leads to a smaller Vainshtein radius, still given
by the same Eq.~\eqref{rV25bis}, but with the charge instead
determined through the lower bound of the accretion condition,
Eq.~\eqref{alpha25-low},
\begin{equation}
r_{V25\text{,local}}^9 \gtrsim \frac{k_5^2 H
r_S^2}{3 |k_2|^3 M^8} =
\left(\frac{3Hr_S}{2}\right)^2 \frac{r_{V25}^9}{\alpha_T}.
\label{rV25accretion-model2}
\end{equation}
The condition determining one or the other scenario of accretion
---~hence the Vainshtein radius~--- can be read off
Eq.~(\ref{g25-slow-acc}). For $\alpha_T/(H r_S)^2\lesssim 1$ the
Vainshtein radius is given by~(\ref{rV25-model2}), while
otherwise we have~(\ref{rV25accretion-model2}). In a given
theory, i.e., a fixed value of $M$, these two Vainshtein radii
coincide for a BH mass such that $\Gamma^{(c)}_\text{acc}\sim H$.
In principle, if a black hole is observed before being fully
quenched, its effective Vainshtein radius could lie between
$r_{V25}$ and the lower bound of $r_{V25\text{,local}}$.

Substituting Eq.~\eqref{rV25-model2} into Eq.~\eqref{Ztt52} we
obtain for the acoustic metric in the small-accretion scenario
\begin{equation}
\label{Z5-model2}
Z_{5(c)}^{tt} \sim \frac{2|k_2|}{9(Hr_S)^3}
\left(\frac{r_S}{r}\right)^3.
\end{equation}
On the other hand when accretion is initially large and undergoes
quenching, the acoustic metric is reduced together with the
Vainshtein radius
\begin{equation}\label{Z5local-model2}
Z^{tt}_{5\text{,local}} \sim
\left(\frac{(3Hr_S)^2}{4\alpha_T}\right)^{1/3} Z^{tt}_{5(c)}.
\end{equation}
Although $Z_5^{tt}$ is proportional to the small quantity
$(r_S/r)^3$, as compared to $(r_S/r)^{3/2}$ for $Z_3^{tt}$ in the
cubic Galileon model, Eq.~\eqref{Z3-model1}, it is enhanced by a
factor $(Hr_S)^{-2}$, therefore the $G_5$ screening is much
stronger than $G_3$. We calculate the observational
consequences of these results in Sec.~\ref{Sec6B}.

\section{Cubic Galileon with a small quintic term}
\label{Sec6}

\subsection{Scalar field solution}
\label{Sec6A}

We now consider the full case of Eqs.~(\ref{G235}), where the
three functions $G_2(X)$, $G_3(X)$ and $G_5(X)$ define the
dynamics of the scalar field. As we will see, this model is
flexible enough to allow us to separate the questions of dark
energy and of black-hole charges. Given the results of the
previous section, we will assume that $k_5$ is much smaller than
$k_2$ and $k_3$ (themselves possibly of order 1). This will not
only avoid too large scalar charges generated by the $1/(M r)^2$
term entering Eq.~(\ref{coefC}), but also allow the model to pass
the known constraint~(\ref{eq:aT0}) on GW speed. More precisely,
we shall assume here that
\begin{equation}
(H r_S)^2 \ll \left|\frac{k_5}{k_3}\right|
\left(\frac{H}{M}\right)^2\ll H r_S,
\label{smallk5}
\end{equation}
which numerically means $10^{-44} \ll |k_5/k_3| (H/M)^2 \ll
10^{-22}$ for the LIGO/Virgo experiments, and \mbox{$10^{-36} \ll
|k_5/k_3| (H/M)^2 \ll 10^{-18}$} for the LISA mission. Although
this may seem a fine-tuned choice, since our theory is part of
Horndeski gravity it enjoys a weakly-broken Galileon
symmetry~\cite{Pirtskhalava:2015nla} which protects the $k_i$
coefficients from large quantum corrections\footnote{A caveat is
that the results of Ref.~\cite{Pirtskhalava:2015nla} were derived
around flat space, which by assumption is not stable in the
theories we consider here.}, which are instead suppressed by the
small ratio $(M/M_\text{Pl})^{2/3} \lesssim 10^{-40}$. Comparing
with the general expression for $r_{V35}$, Eq.~\eqref{rV35}, we
see that the left inequality is equivalent to $r_S\ll r_{V35}$,
i.e., the requirement that this black hole be surrounded by a
$G_5$-dominated region. Otherwise, the $G_5$ part of the solution
is never relevant for this black hole.

With the above assumptions, Eq.~(\ref{J0}) imposes
\begin{equation}
\dot\varphi_c = \frac{k_2 M^2}{3 k_3 H}
\left[1
+\mathcal{O}\left(\frac{k_5 H^2}{k_3 M^2}\right)\right],
\label{phiDotG2G3G5}
\end{equation}
which is thus almost equal to Eq.~(\ref{phiDotG2G3}), up to a
fully negligible relative correction, much smaller than $H r_S
\sim 10^{-18}$ for LISA black holes. Thus the energy density of
the background under these assumptions is essentially the same as
in the $G_2$-$G_3$ model, Eq.~\eqref{eq:rhoG2G3},
\begin{equation}\label{eq:rhoG2G3bis}
\rho_\varphi = \frac{|k_2|^3M^4}{9k_3^2 H^2}
\left[1+\mathcal{O}\left(\frac{k_5 H^2}{k_3 M^2}\right)\right].
\end{equation}
On the other hand, the speed of GWs on the cosmological
background is corrected by
\begin{equation}
\alpha_T
\approx 2\left(\frac{k_2}{3 k_3}\right)^3 k_5
\left(\frac{M}{H}\right)^2.
\label{eq:alphaT-model3}
\end{equation}
Note that $\alpha_T$ scales as $H^{-2}$ here, compared to
$H^{-8}$ for the $G_2$-$G_5$ model, while the energy density
scales as $H^{-2}$ (vs $H^{-6}$) ---~the ratio
$\alpha_T/\rho_\varphi$ is constant in the present model. With
these expressions we can also rewrite our condition for the
validity of our perturbative expansion~\eqref{smallk5} as
\begin{equation}\label{eq:smallk5-physical}
(Hr_S)^2 \ll \frac{3H^2}{2\rho_\varphi} |\alpha_T| \ll {Hr_S},
\end{equation}
which confirms that whenever our expansion is valid, $|\alpha_T|
\ll 10^{-18}$ for LISA black holes, well within the
constraint~\eqref{eq:aT0} even for full self-acceleration. The
small quintic contribution will only have local consequences, at
distances of order of the Schwarzschild radius $r_S$. We shall
see that it actually imposes the scalar charge
$\alpha_\text{BH}$.

Since the perturbative method of Sec.~\ref{Sec4B} relies on
expansions in powers of the small dimensionless quantity $H r_S$,
we must however be careful when assuming that $k_5$ is small, as
ratios of small parameters might be of any size. A convenient
technique is to define a new parameter
\begin{equation}
\overline k_5 \equiv \frac{k_5}{(H r_S)^2},
\label{k5Bar}
\end{equation}
of ``reasonable'' size. Since this can be visually useful to
understand the order of magnitude of the various terms we will
write below, let us copy our assumptions~(\ref{smallk5}) in terms
of this new notation,
\begin{equation}
1 \ll \left|\frac{\overline k_5}{k_3}\right|
\left(\frac{H}{M}\right)^2
\ll \frac{1}{H r_S} \sim 10^{22},
\label{smallk5Bar}
\end{equation}
(or $\sim 10^{18}$ for the LISA mission). When $|k_3| \sim M/H
\sim 1$, the parameter $|\overline k_5|$ is thus assumed to be
negligible with respect to $1/(H r_S)$, so that one can perform
safely our expansions in powers of $H r_S$. And when both $k_3$
and $\overline k_5$ occur at the same order, we may neglect the
former, because of the first inequality in~(\ref{smallk5Bar}). We
did check that these assumptions are consistent with our results
below at each step of our calculations.

Because of them, the large-distance behavior of the solution is
extremely close to that of Sec.~\ref{Sec4}, and we notably
recover Eqs.~(\ref{phiDotRootConsistent}) for a double root of
the discriminant $\Delta$ to exist near $r\approx 1/(\sqrt{3}H)$.
On the other hand, the quintic term plays a dominant role at
small distances, and we find that the existence of a local double
root of $\Delta$ imposes
\begin{subequations}
\label{eqs235}
\begin{eqnarray}
\dot\varphi_\text{BH} &=& \frac{k_2 M^2}{3 k_3 H}
\biggl[
1 + \Bigl(
\frac{4 \overline k_5 H^2}{3 \sqrt{3}\, k_3 M^2}
+ 6 \sqrt{3}
\nonumber\\
&&- \frac{3^6 \sqrt{3}\, k_3 M^2}{2^4
\overline k_5 H^2}\Bigr) H r_S
+\mathcal{O}\left(H^2 r_S^2\right)
\biggr],
\label{phiDot235}\\
r_\text{root}^\text{close} &=&
\frac{3}{2} r_S\left[1+
\frac{3^4 \sqrt{3}\,
|k_3| M^2 r_S}{2^5 \sqrt{5}\, \overline k_5 H}
+ \mathcal{O}\left(H^2 r_S^2\right)
\right],
\label{root235}\\
r_\text{root}^\text{far} &=&
\frac{1}{\sqrt{3} H}
+ \left(\frac{2^4 \overline k_5 H^2}{3^2 k_3 M^2}
+\mathcal{O}(1)\right)r_S
+ \mathcal{O}\left(H r_S^2 \right),
\qquad\null
\label{rootFar235}\\
\alpha_\text{BH} &=& 2^3 \left(\frac{k_2}{9 k_3}\right)^2
\overline k_5
\biggl[
1 + \frac{3^3 k_3 M^2}{2 \overline k_5 H^2}
+\mathcal{O}\left(\frac{k_3^2}{\overline k_5^2}\right)
\nonumber\\
&& + \frac{\left(2^6 \overline k_5^2 H^4
+ 2^5 3^3 k_3 \overline k_5 H^2 M^2
- 3^8 k_3^2 M^4\right)^2}{2^9 3 \sqrt{3}\,
k_3 \overline k_5^3 H^6 M^2} H r_S
\nonumber\\
&&+\mathcal{O}\left(H^2 r_S^2\right)\biggr].
\label{alphaBH235}
\end{eqnarray}
\end{subequations}
As before, we only display the first two terms of these
expansions, although we did compute higher orders. We also
display the large-distance double root $r_\text{root}^\text{far}$
in Eq.~(\ref{rootFar235}), obtained from the replacement
of~(\ref{alphaBH235}) into~(\ref{rootConsistent}), to parallel
the presentation of Eqs.~(\ref{correct23}) for the cubic Galileon
of Sec.~{\ref{Sec4}. These intricate expressions show that such a
case of $G_2$, $G_3$ and $G_5$ together is highly non-trivial. In
particular, the existence of simultaneous ratios $\overline
k_5/k_3$ and $k_3/\overline k_5$ underlines that fully neglecting
one of these two parameters would be inconsistent.\footnote{It is
interesting to note that when the left-hand side of our
hypothesis~(\ref{smallk5}) or (\ref{smallk5Bar}) is not
satisfied, then the first correction in Eq.~(\ref{alphaBH235})
behaves as $\left(2 k_2 M/H\right)^2/(3 k_3)$, i.e., very similar
to what we found in Eq.~(\ref{correctAlpha}) for the pure
quadratic plus cubic Galileon model, up to a factor $4$. This
confirms that when $|k_5/k_3|$ is even smaller than $(H r_S)^2$,
the present model progressively behaves as this cubic Galileon of
Sec.~\ref{Sec4}. The next correction in~(\ref{alphaBH235}),
proportional to $1/\overline k_5$, has a meaning only when our
assumption~(\ref{smallk5Bar}) is satisfied, on the other hand.}
But the $1$ starting all square brackets indeed dominate over the
next terms, under our assumptions~(\ref{smallk5Bar}), and the
global factors of such square brackets thus give the lowest-order
values. We can therefore conclude that in the present model, it
is possible to assume a linear time dependence~(\ref{linearTime})
of the scalar field everywhere. The two double roots of the
discriminant $\Delta$ are indeed consistent with each other if
the time derivative (\ref{phiDot235}) only slightly differs from
its cosmological background~(\ref{phiDotG2G3G5}). This confirms
that the present model behaves as the (quadratic plus) cubic one
of Sec.~\ref{Sec4} at large distances. On the other hand,
Eq.~(\ref{root235}) shows that the location of the double root of
$\Delta$ at small distances almost equals the one we found in
(\ref{rootc25}) for the quintic Galileon of Sec.~\ref{Sec5}.
Therefore, the $G_5$ term dominates at small distances, as
expected from its $1/(M r)^2$ contribution to Eq.~(\ref{coefC}).
The predicted scalar charge~(\ref{alphaBH235}) reads at lowest
order
\begin{equation}
\alpha_\text{BH} \approx
8 \left(\frac{k_2}{9 k_3 H r_S}\right)^2 k_5,
\label{alphaBH235approx}
\end{equation}
where we have now replaced the intermediate
notation~(\ref{k5Bar}) by its actual expression in terms of the
theory parameter $k_5$. We can express this in terms of the
physical parameters of the cosmological background of this model
as
\begin{equation}
\alpha_\text{BH} \approx \left( \frac{2}{3Hr_S} \right)^2
\sqrt{\frac{-k_2}{3}}
\frac{\alpha_{T}}{\sqrt{\rho_\varphi/3H^2}}.
\end{equation}
This is to be compared to Eq.~(\ref{alphaBHc25}) that we found
for the quintic Galileon of Sec.~\ref{Sec5} ---~up to a factor of
$\sqrt{8}$, it can be shown to be the same expression when
translated into the physical parameters, despite the different
proportionality with respect to $k_5$. For this general model, we
are free to adjust $\rho_\varphi$ independently of $k_5$ (or
$\alpha_T$) and therefore the charge can be small even in the
case of full self-acceleration.
In other words, even if we assume $|k_2|\sim |k_3|\sim M/H
\sim 1$, the scalar charges of black holes are no longer required
to be huge, contrary to Sec.~\ref{Sec5}.

Let us also underline that contrary to Eqs.~(\ref{correctAlpha})
and (\ref{alphaBHc25}) of the previous sections,
$\alpha_\text{BH}$ is independent of the theory parameter $M$ in
the present model. This means that when the Galileon is not fully
self-accelerating (i.e., $M/H$ is small if we choose $|k_2|\sim
|k_3| \sim 1$), the scalar charge~(\ref{alphaBH235approx}) does
not change. It is quite surprising that the consistency of the
present solution imposes such a large scalar charge even when the
scalar field has actually a negligible influence in cosmology.

\subsection{Accretion}
\label{Sec6Accretion}

The effect of accretion may already be estimated from
Eqs.~\eqref{phiDot235} and \eqref{alphaBH235}. Its
rate~(\ref{Gammaacc}) here evaluates to
\begin{eqnarray}
\Gamma^{(c)}_\text{acc} = |\dot\varphi_c\,
\alpha_\text{BH}|
&\approx&
\frac{2^3 |k_2|^3}{3^5 |k_3|^3 H^3 r_S^2}\, |k_5| M^2
\nonumber\\
&\approx&
\left(\frac{2}{3Hr_S}\right)^2
|\alpha_T|\, H,
\label{eq:Gamma-model3}
\end{eqnarray}
which is the same expression as for the quintic Galileon
Eq.~\eqref{g25-slow-acc}: It is strongly enhanced by the
$1/r_S^2$ dependence of the scalar charge $\alpha_\text{BH}$.
However, here we can vary $\alpha_T$ independently of the
Galileon energy density, by reducing $k_5$,
requiring\footnote{When combining the limit of
Eq.~(\ref{restriction}), $|k_5| (M/H)^2 \sim (H r_S)^2$, with our
assumptions~(\ref{smallk5}), this implies $(H r_S)^2\ll |k_5|\ll
(H r_S)^{3/2}$ with $1 \gtrsim M/H \gtrsim (H r_S)^{1/4}$.}
\begin{equation}
|k_5| \left(\frac{M}{H}\right)^2
\lesssim (H r_S)^2,
\label{restriction}
\end{equation}
when setting $|k_2| \sim |k_3| \sim 1$ by redefining the
variables entering functions~(\ref{G235}).

In fact, we can rewrite our condition \eqref{smallk5} as
\begin{equation}
\label{eq:G5G3-accr-validity}
1 \ll \frac{\Gamma_\text{acc}^{(c)}/H}{\rho_\varphi/(3H^2)}
\ll (Hr_S)^{-1}.
\end{equation}
As we have already mentioned after Eq.~\eqref{smallk5}, when the
left inequality is not satisfied, $r_S > r_{V35}$ and there is no
$G_5$-dominated region around this black hole. For large enough
black holes, we would instead recover the cubic Galileon
behavior already studied in section \ref{Sec4}, in which
accretion is always small. If there is a $G_5$ dominated region
at all, then expression~\eqref{eq:G5G3-accr-validity} means that
a black hole can only slowly accrete on the cosmological
background, $\Gamma_\text{acc}^{(c)}<H$ when the Galileon is not
driving the acceleration, $\rho_\varphi < 3H^2$. In the full
self-acceleration case, all black holes slowly accreting on the
cosmological background are in $G_3$ domination.

It is also worth reiterating that the middle term in
Eq.~\eqref{eq:G5G3-accr-validity} is constant (see
Eqs.~\eqref{eq:rhoG2G3bis} and \eqref{eq:alphaT-model3}), so if
the black hole is in the $G_5$-dominated accretion regime now, it
was so in the past. This directly results from the constancy of
$r_{V35}$, Eq.~\eqref{rV35}. In any case, LIGO/Virgo black holes
are slowly accreting in this model only if $|\alpha_T|\lesssim
10^{-43}$, which would pass the constraint~\eqref{eq:aT0} by 28
orders of magnitude and be safely quasi-stationary whenever
conditions~\eqref{eq:G5G3-accr-validity} are satisfied.
\medskip

Just as in the case of the simple quintic Galileon, there are
always small enough black holes, $(3Hr_S/2)^2< \alpha_T$, which
have a large accretion rate on the cosmological background
$\dot\varphi_c$. These black holes undergo the quenched accretion
scenario and absorb the energy density stored in the scalar,
decreasing their charge. The expressions for the configuration
once the accretion is quenched are just as in the quintic model
of section~\ref{Sec5}, i.e., we have
\begin{equation}
\label{alphaLocal235}
\alpha_\text{BH} \approx 2 k_5
\left(\frac{2 \dot\varphi_\text{local}}{3 M^2 r_S}\right)^2,
\end{equation}
cf.~Eq.~\eqref{alphaLocal}, and requiring
$\Gamma_\text{acc}=|\alpha_\text{BH}\, \dot\varphi_\text{local}|
\gtrsim H$ gives again
\begin{subequations}
\begin{align}
|\dot\varphi_\text{local}| &\gtrsim
\left(\frac{9 H M^4 r_S^2}{8 |k_5|}\right)^{1/3}
\label{phi235-low}\\
|\alpha_\text{BH}| &\gtrsim
2\left(\frac{|k_5| H^2}{9 M^4 r_S^2}\right)^{1/3},
\label{alpha235-low}
\end{align}
\end{subequations}
cf.~Eqs.~\eqref{phi25-low} and \eqref{alpha25-low}. Again, this
is a conservative lower bound for the charge in this alternative
scenario. The above bounds are identical for both the models
since these black holes decouple from the cosmological background
produced by the $G_3$ term. The black-hole charge is determined
by the fact the object is in $G_5$ domination and it is the $G_5$
term, shared by both the present model and the simple quintic
Galileon, that sets all the relevant properties of the quenched
quasi-stationary configuration.
The contribution of the energy-momentum tensor of the scalar
field close to the black hole is thus the same as in
Sec.~\ref{Sec5LargeAccretion}, i.e., negligible with respect to
the assumed $r_S$ in metric~(\ref{metric}), whose form can thus
be trusted.\footnote{Curiously enough, in the present
model, this scalar's energy is \textit{also} negligible during
the process of fast scalar accretion, even when the scalar charge
takes its largest value~(\ref{alphaBH235approx}). This is due to
our assumption of a small parameter $|k_5|$ in
Eq.~(\ref{smallk5}). One indeed finds that the scalar's energy
contributes at order $\sim (M/H)^2 |k_5|/H \ll (M/H)^4 r_S$,
therefore negligible with respect to $r_S$.}

Screening does not change any of this discussion, since the
unscreened charge $\alpha_\text{BH}$ is responsible for the
energy flux~(\ref{energy-flux}) into the horizon. Nevertheless,
screening is still important for the observable effects, as it
does suppress the scalar-wave emission. We turn to this question
now.

\subsection{Vainshtein screening}
\label{Sec5C}
In this general model, we have the possibility of the full set of
Vainshtein screened regimes, as illustrated in Fig~\ref{Fig1}.

In the setup with small accretion, we obtain the Vainshtein radii
{}from the general expressions Eqs.~(\ref{rV35})--(\ref{rV23})
using Eq.~\eqref{alphaBH235approx} for the black-hole charge. The
smallest one is independent of the black hole charge and of time,
\begin{equation}
r_{V35}^3 = \frac{|k_5| r_S}{2 |k_3| M^2}
= \frac{3|\alpha_T|}{4\rho_\varphi} r_S.
\label{rV35-model3}
\end{equation}
where we have again expressed the model parameters in terms of
the combination of the physical properties of the cosmological
background. We repeat here that the lower limit of
condition~\eqref{smallk5} is equivalent to $r_S \ll r_{V35}$,
i.e., $G_5$-domination on the scales of $r_S$. The intermediate
Vainshtein radius is given by
\begin{equation}
r_{V25}^3 \approx \frac{2|k_5|}{9 |k_3| H M^2}
= \frac{4}{9}\frac{r_{V35}^3}{H r_S}.
\label{rV25bis-model3}
\end{equation}
It corresponds to what we found in Eq.~(\ref{rV25-model2}) of
Sec.~\ref{Sec5}, but the difference is that $\alpha_\text{BH}$
now takes the value (\ref{alphaBH235approx}) instead of
(\ref{alphaBHc25}). Finally, a third and largest Vainshtein
radius is given by
\begin{equation}
r_{V23}^3 \approx
\frac{8}{(9 H M)^2}
\frac{|k_5|}{|k_3| r_S} = \frac{4}{9}\frac{r^3_{V25}}{H r_S}.
\label{rV23-model3}
\end{equation}
The hierarchy of the Vainshtein radii is determined by the black
hole size $Hr_S$. Contrary to Eqs.~(\ref{rV23cubic}) and
(\ref{rV25-model2}) of the previous sections, the three
Vainshtein radii~(\ref{rV35-model3})--(\ref{rV23-model3}) here
depend on the theory parameter $M$, and more precisely, they are
all proportional to $1/M^2$. This is due to the fact that the
scalar charge~(\ref{alphaBH235approx}) is now independent of $M$.
When self-acceleration is not full, these three transition radii
become larger, and the Vainshtein screening is more efficient.

In the quenched-accretion scenario, we need to instead use the
bound on the quenched charge~\eqref{alpha235-low} to obtain:
\begin{equation}
r_{V25,\text{local}}^9 \gtrsim
\frac{k_5^2 Hr_S^2}{3\, |k_2|^3 M^8} =
\left(\frac{3 H r_S}{2}\right)^2 \frac{r_{V25}^9}{|\alpha_T|},
\label{rV25bis-model3-local}
\end{equation}
recovering expression Eq.~\eqref{rV25accretion-model2} for the
simple quintic model, while the largest Vainshtein radius
becomes,
\begin{equation}
r_{V23\text{,local}}^9\gtrsim
\frac{8\,|k_3^3 k_5|}{9\, k_2^6 M^{10}} H^2 r_S =
\left(\frac{3 H r_S}{2}\right)^4 \frac{r_{V23}^9}{\alpha_T^2},
\label{rV23bis-model3-local}
\end{equation}
where we have used the expression for $\alpha_T$ on the
cosmological background of this model,
Eq.~\eqref{eq:alphaT-model3}, to relate the Vainshtein radii of
the local quenched solutions to the radii~(\ref{rV25bis-model3})
and (\ref{rV23-model3}). Equation~\eqref{eq:Gamma-model3} implies
that large accretion occurs whenever $(Hr_S)^2/|\alpha_T| < 1$.
Thus the local Vainshtein radii are always smaller than the
respective cosmological ones, as Eqs.~(\ref{rV25bis}) and
(\ref{rV23}) also show because the scalar charge
$|\alpha_\text{BH}|$ is reduced. Using the upper limit of our
assumptions~(\ref{smallk5}), one can also prove that the order
$r_{V35} < r_{V25,\text{local}} < r_{V23,\text{local}}$ is
maintained.

The important criterion to note is the relative size of $r_{V35}$
versus the wavelength of the emitted gravitational waves,
$\lambda\sim300r_S$. If $r_{V35}\gg \lambda$, then the emission
of gravitational waves will be governed by the $G_5$ term in the
acoustic metric $Z^{\mu\nu}$. Otherwise, $r_{V35}<\lambda$ and
the emission is governed by the $G_3$ term, even if the black
hole itself is in $G_5$ domination, $r_{V35}>r_S$.

The effect of the screening in the end boils down to the
normalization of the effective metric at the scale $\lambda$.
Taking the expression~\eqref{Ztt52} and re-expressing $r_{V25}$
using Eq.~\eqref{rV25bis-model3} gives
\begin{equation}
Z_{5(c)}^{tt} \sim \frac{1}{Hr_S}
\left(\frac{r_{V35}}{\lambda}\right)^3 |k_2|.
\label{Z5c-model3}
\end{equation}

For the quenched accretion scenario, we can use
Eq.~\eqref{rV25bis-model3-local} to obtain
\begin{equation}
Z_{5\text{,local}}^{tt} \sim
\left[\frac{(3H r_S)^2}{4|\alpha_T|}\right]^{1/3} Z_{5(c)}^{tt},
\end{equation}
which is reduced compared to $Z_{5(c)}^{tt}$.

Finally when the scale $\lambda$ is in a $G_3$-domination region,
$r_{V35}<\lambda$, but the black hole itself is in $G_5$
domination, using Eqs.~\eqref{rV23-model3} and
\eqref{rV23bis-model3-local} in Eq.~\eqref{eq:Z3-gen}, we obtain
\begin{align}
&Z_{3(c)}^{tt} \sim \frac{1}{Hr_S}
\left(\frac{r_{V35}}{\lambda}\right)^{3/2} |k_2| \\
&Z_{3\text{,local}}^{tt} \sim
\left[\frac{(3H r_S)^2}{4|\alpha_T|}\right]^{1/3} Z_{3(c)}^{tt}
\end{align}
and we recover that the $G_3$ and $G_5$ contributions to the
total $Z^{\mu\nu}$ are of the same magnitude at $r=r_{V35}$ and
they have a common time dependence.

Again, the overall radiation flux will depend on the combination
of the metric normalization $Z^{tt}$ and the black hole charges.
We discuss the potential for observability for this model in
section~\ref{Sec6C}.

\section{Observational consequences}
\label{Observations}

\subsection{Radiation from binary inspirals}
The most direct observational constraint on these black-hole
charges results from an additional channel for radiation in
black-hole mergers, allowing them to proceed more quickly than in
general relativity.

In a binary system of two bodies $A$ and $B$ at equilibrium, with
masses of the same order of magnitude, $m_A \sim m_B \sim
r_S/(2G)$, and a negligible eccentricity, the energy flux carried
away by gravitational radiation is dominated in general
relativity by the quadrupole term
\begin{equation}
F_\text{GR} \approx
\frac{2}{5 G} \left(\frac{r_S}{r_{AB}}\right)^5,
\label{FGR}
\end{equation}
where $r_{AB}$ denotes the interbody distance. This is the
simplest writing, but it is useful to reexpress it in terms of
the orbital angular frequency $\Omega_\text{p} \equiv 2\pi/P$,
where $P$ is the orbital period. This is achieved thanks to
Kepler's third law (at its lowest, Newtonian, order),
$\Omega_\text{p}^2 r_{AB}^3 = G (m_A+m_B) \approx r_S$, which
implies
\begin{equation}
\frac{r_S}{r_{AB}} \approx
\left(\Omega_\text{p} r_S\right)^{2/3}.
\label{rsOnrAB}
\end{equation}
Therefore Eq.~(\ref{FGR}) is proportional to
$\left(\Omega_\text{p} r_S\right)^{10/3}$. Twice the orbital
frequency is the wave frequency, which is directly observable,
while the chirp mass is of order $r_S$.

In scalar-tensor theories, the binary also emits scalar waves,
whose dominant contributions to the energy flux
read~\cite{Damour:1992we}
\begin{subequations}
\label{FscalarFull}
\begin{eqnarray}
F_\text{scalar} &=&
\frac{F_\text{scalar}^\text{dipole}}{z_{\lambda,1}}
+ \frac{F_\text{scalar}^\text{quadrupole}}{z_{\lambda,2}},
\label{Fscalar}\\
F_\text{scalar}^\text{dipole} &\approx&
\frac{1}{48 G |k_2|} \left(\frac{r_S}{r_{AB}}\right)^4
\left(\alpha_A^\text{eff}-\alpha_B^\text{eff}\right)^2,
\label{FDipole}\\
F_\text{scalar}^\text{quadrupole} &\approx& \frac{1}{15 G |k_2|}
\left(\frac{r_S}{r_{AB}}\right)^5
\alpha_A^\text{eff} \alpha_B^\text{eff}.
\label{FQuadrupole}
\end{eqnarray}
\end{subequations}
The coefficients $z_{\lambda,\ell}$ are the Vainshtein screening
factors discussed in Sec.~\ref{VainshteinScreening}, that we
shall further describe below. The quadrupolar
term~(\ref{FQuadrupole}) is of the same order of magnitude as the
general relativistic prediction~(\ref{FGR}), but multiplied by
the square of the dimensionless scalar charge $\alpha_\text{BH}
\sim \alpha_A^\text{eff} \sim \alpha_B^\text{eff}$, which is of
order $1$ in the cubic Galileon model of Sec.~\ref{Sec4} but may
be large in presence of the quintic term, as seen in
Secs.~\ref{Sec5} and~\ref{Sec6}. On the other hand, the dipolar
term~(\ref{FDipole}) is generically larger than the GR
quadrupole~(\ref{FGR}), because it involves a smaller power of
$r_S/r_{AB} = 2 G m/(r_{AB} c^2)$, i.e., it is of a
\textit{lower} post-Newtonian order. However, note that it needs
the two scalar charges $\alpha_A^\text{eff}$ and
$\alpha_B^\text{eff}$ to differ. Although we assume that the two
black holes have similar masses $m_A \sim m_B$, this means that
they must not be strictly equal for this dipolar
term~(\ref{FDipole}) to be significant. Moreover, even when
$m_A\neq m_B$, the cubic Galileon model of Sec.~\ref{Sec4}
predicted that $\alpha_\text{BH}$, Eq.~(\ref{correctAlpha}), does
not depend on the black hole mass (up to negligible relative
corrections of order $H r_S$). Therefore, although the
dipole~(\ref{FDipole}) is generically the dominant scalar
contribution to the energy flux, it happens to be negligible in
the case of the cubic Galileon model.

In Eqs.~(\ref{FscalarFull}), the superscripts ``eff'' of
$\alpha_A^\text{eff}$ and $\alpha_B^\text{eff}$ come from the
fact that our test scalar field solution~(\ref{scalarsolution})
depends on the discriminant $\Delta$, Eq.~(\ref{discriminant}),
which involves the combination $(C-\alpha_\text{BH}r_S/r^2)$ and
not only the scalar charge $\alpha_\text{BH} r_S$. This means
that the actual black hole-scalar coupling strength is not merely
$\alpha_\text{BH}$, but rather the coefficient of the $-r_S/r^2$
main term of this combination, namely
\begin{equation}
C-\frac{\alpha_\text{BH}r_S}{r^2} =
-\left(\frac{k_3 \dot\varphi_\text{BH}^2}{M^2}
+\alpha_\text{BH}\right)\frac{r_S}{r^2}
+\mathcal{O}\left(\frac{r_S^2}{r^3}\right).
\label{couplingConstant}
\end{equation}
The coupling constants entering Eqs.~(\ref{FscalarFull}),
for each body $A$ and $B$, are thus given by
\begin{equation}
\alpha_\text{BH}^\text{eff} \equiv
\alpha_\text{BH} +
\frac{k_3 \dot\varphi_\text{BH}^2}{M^2}.
\label{correctAlphaeff}
\end{equation}
This is the black-hole analogue of the effective matter-scalar
coupling strength $\alpha_\text{eff}$ defined in Eq.~(10) of
Ref.~\cite{Babichev:2012re}. In the cubic Galileon model of
Sec.~\ref{Sec4}, Eqs.~(\ref{phiDotG2G3}), (\ref{correctPhiDot})
and (\ref{correctAlpha}) imply that this effective scalar charge
reads
\begin{equation}
\alpha_\text{BH}^\text{eff} = \frac{4}{9 k_3}
\left(\frac{k_2 M}{H}\right)^2
+\mathcal{O}\left(H r_S\right)
=\frac{4}{3}\alpha_\text{BH}+\mathcal{O}\left(H r_S\right),
\label{correctAlphaeffCubic}
\end{equation}
which is of the same order of magnitude as $\alpha_\text{BH}$. In
the quintic Galileon model of Sec.~\ref{Sec5}, the quantity $C$,
Eq.~(\ref{coefC}), does not involve any correction proportional
to $1/r^2$, therefore $\alpha_\text{BH}^\text{eff} =
\alpha_\text{BH}$. Finally, in the full (quadratic plus cubic
plus quintic) model of Sec.~\ref{Sec6}, the $k_3
\dot\varphi_\text{BH}^2/M^2$ correction
entering~(\ref{correctAlphaeff}) does not vanish but is
negligibly small because of our assumptions~(\ref{smallk5Bar}),
\begin{equation}
\left|\frac{k_3
\dot\varphi_\text{BH}^2}{M^2\alpha_\text{BH}}\right|
\approx \left(\frac{3M}{2H}\right)^2
\left|\frac{k_3}{2 \overline k_5}\right| \ll 1,
\end{equation}
therefore $\alpha_\text{BH}^\text{eff} \approx \alpha_\text{BH}$.

The crucial difference with standard scalar-tensor
theories~\cite{Damour:1992we} (i.e., with a standard kinetic term
(\ref{G2}) alone) is that there exists a Vainshtein screening in
the present nonlinear Galileon theories, which grossly changes
the normalization of the scalar fluctuations, as described in
Sec.~\ref{VainshteinScreening}: Scalar perturbations and
interactions via scalar exchange behave as if each of the scalar
charges were renormalized by a small factor $z_\lambda^{-1/2}$.
This ensures that the effect of the scalar wave emission on
the binary dynamics remains perturbative with respect to GR and
can be described by Eqs.~\eqref{FscalarFull} evaluated on the
unmodified post-Newtonian trajectories. For estimating the scalar
force between the bodies, the relevant scale $\lambda$ is the
inter-body distance $r_{AB}$. However, the emission of
gravitational waves (including helicity-0 ones due to the scalar
field) is a collective phenomenon which builds up at the scale of
the gravitational wavelength, therefore the scale $\lambda$ to
insert in the reduction factor is rather this wavelength. This is
what was carefully derived in Eq.~(A.42) of
Ref.~\cite{Dar:2018dra} for the case of the cubic Galileon in the
usual static, asymptotically flat configuration, rather than the
cosmological boundary being considered here. We will adapt these
results for our case, under the assumption that, deep in the
Vainshtein region, the effect of the cosmological boundary is
limited to inducing the scalar charges (as confirmed below
Eq.~(34) of Ref.~\cite{Babichev:2012re} for $r \gg r_S$, where
the same Vainshtein screening factor was found for both of these
asymptotic cases). This is not strictly correct, since the
presence of $\dot\varphi_c$ gives rise to mixed $tr$ terms in the
acoustic metric, and therefore a tilting of the sound cone with
respect to the light cone. We are only looking for
order-of-magnitude estimates, so we will not study this detail in
the present work. Reference~\cite{Dar:2018dra} proved that the
Vainshtein reduction factor reads
\begin{equation}
z_{\lambda,\ell}^\text{cubic} \approx
\frac{1}{4} \left(\Omega_\text{p} r_{AB}\right)^{3-\ell}
\left(\Omega_\text{p} r_{V23}\right)^{3/2}.
\label{zcubic}
\end{equation}
The large factor $\left(\Omega_\text{p} r_{V23}\right)^{3/2}$ is
the expected one from evaluating the Vainshtein screening at the
distance of a few ($\pi$) wavelengths. The extra factor
$\left(\Omega_\text{p} r_{AB}\right)^{3-\ell}$ is more subtle, as
it depends on the ratio of orthoradial to radial velocities of
scalar perturbations. For $\ell = 2$ (quadrupole), it is of order
$0.2$ for LIGO/Virgo binaries, and $0.1$ for LISA. We shall take
it into account below when the binaries are in the
$G_3$-dominated region, but they do not change significantly our
order-of-magnitude estimates.

The same careful analysis has not been performed for the quintic
Galileon, as there is no known covariant way to diagonalize the
kinetic terms of the spin-2 and spin-0 degrees of freedom. But
aside from some possible factors of $\left(\Omega_\text{p}
r_{AB}\right)$ which do not change much the orders of magnitude
(and which would increase the predicted scalar effects, therefore
it is conservative not to include them), the above results
confirm that the Vainshtein screening factor should be evaluated
at a few wavelengths. From our discussion of
Sec.~\ref{VainshteinScreening}, we may thus estimate that in a
$G_5$-dominated region, one should use
\begin{equation}
z_{\lambda}^\text{quintic} \sim
\left(\Omega_\text{p} r_{V25}\right)^{3}.
\label{zquintic}
\end{equation}
Note two crucial differences with respect to Eq.~(\ref{zcubic}):
The exponent is now $3$ instead of $\frac{3}{2}$, and the
Vainshtein radius entering this expression is $r_{V25}$ instead
of $r_{V23}$.

Before computing the scalar effects for our three models of
Secs.~\ref{Sec4}, \ref{Sec5} and~\ref{Sec6}, let us quote the
numerical values we shall use. The bandwidth of LIGO and Virgo
interferometers is between $30$~Hz and $10^3$~Hz, but this is for
the lowest frequencies that they accumulate the largest number of
cycles of their inspiral phase, and can therefore significantly
constrain deviations from general relativity. We shall thus take
$\Omega_\text{p} = \pi \nu$, with $\nu \approx 30$~Hz (note the
factor $\pi$ instead of $2\pi$ because the GW frequency is twice
that of the orbit). In the case of LISA, we shall similarly use
the lowest frequency of its bandwidth, namely $\nu \approx
10^{-4}$~Hz. This gives
\begin{subequations}
\begin{eqnarray}
\Omega_\text{p} &\approx& 10^2
\text{~rad s}^{-1} \text{ for LIGO/Virgo},\\
\Omega_\text{p} &\approx& 3 \times 10^{-4}
\text{~rad s}^{-1} \text{ for LISA}.
\end{eqnarray}
\end{subequations}
The largest number of observed cycles correspond to rather light
black holes, i.e., of about $10\, m_\odot$ in the case of
LIGO/Virgo, and $10^5\, m_\odot$ for LISA. This corresponds to
\begin{subequations}
\label{Omegaprs}
\begin{eqnarray}
\Omega_\text{p} r_S &\approx& 10^{-2}\text{ for LIGO/Virgo},\\
\Omega_\text{p} r_S &\approx& 3 \times 10^{-4}\text{ for LISA},
\end{eqnarray}
\end{subequations}
and
\begin{subequations}
\label{Hrs}
\begin{eqnarray}
H r_S &\approx& 2 \times 10^{-22}\text{ for LIGO/Virgo},\\
H r_S &\approx& 2 \times 10^{-18}\text{ for LISA}.
\end{eqnarray}
\end{subequations}
In LIGO/Virgo tests of general relativity, the parameter denoted
as $\varphi_{-2}$ in Ref.~\cite{LIGOScientific:2021sio}
quantifies the allowed correction to dipolar radiation. This
reference's Figure~6 gives the constraint $\varphi_{-2} <
10^{-3}$. The ratio of the dipolar term in Eq.~(\ref{Fscalar}) to
the GR prediction (\ref{FGR}) is thus constrained by
\begin{equation}
\frac{F_\text{scalar}^\text{dipole}/z_{\lambda,1}}{F_\text{GR}}
< 10^{-3}\text{ with LIGO/Virgo},
\label{ratioDipole}
\end{equation}
with $z_{\lambda,1}$ given by Eq.~(\ref{zcubic}) if the
wavelength lies in the $G_3$-dominated region, or by
Eq.~(\ref{zquintic}) if it lies in the $G_5$-dominated region.
In the cubic Galileon model of Sec.~\ref{Sec4}, all black holes
have the same $\alpha_\text{BH}$ (up to negligible corrections),
therefore the dipole vanishes, and we may use the constraints on
the parameter $\varphi_0$ of this same
Ref.~\cite{LIGOScientific:2021sio}, which quantifies the allowed
correction to quadrupolar radiation. Its Figure~6 gives
$\varphi_0 < 5\times 10^{-2}$, implying
\begin{equation}
\frac{F_\text{scalar}^\text{quadrupole}/
z_{\lambda,2}}{F_\text{GR}}
< 5\times 10^{-2}\text{ with LIGO/Virgo}.
\label{ratioQuadrupole}
\end{equation}
With the future LISA mission, one can expect to observe about
30000 cycles for black hole masses of order $10^5
m_\odot$~\cite{Blanchet:2023bwj}, with a large signal-to-noise
ratio of a few hundreds. The deviations from GR should thus be
tested at least at the level
\begin{equation}
\frac{F_\text{scalar}}{F_\text{GR}} \lesssim
\frac{1}{100\times 30000} \approx
3\times 10^{-7}\text{ with LISA},
\label{ratioLISA}
\end{equation}
if the detected GWs are consistent with the general relativistic
templates. Significantly tighter bounds are actually predicted in
Ref.~\cite{Perkins:2020tra}, for various populations of BHs and
combined experiments (see notably its Fig.~11), therefore the
above bound~(\ref{ratioLISA}) is conservative.

\subsection{Cubic Galileon}
\label{Sec6Aobs}

In the cubic Galileon model of Sec.~\ref{Sec4} (i.e., with
$k_5 = 0$), the Vainshtein radius~(\ref{rV23cubic}) gives
\begin{subequations}
\label{OmegaprV23}
\begin{eqnarray}
\Omega_\text{p} r_{V23} &\approx& 2 \times 10^{12}
\text{ for LIGO/Virgo},\\
\Omega_\text{p} r_{V23} &\approx& 10^8\text{ for LISA}.
\end{eqnarray}
\end{subequations}
These large values mean that the binaries are deep within the
screened region where $G_3$ dominates.

We underlined above that scalar dipolar radiation is vanishingly
small in the present cubic Galileon model, because all black
holes have the same dimensionless scalar charge
$\alpha_\text{BH}$, see~(\ref{correctAlpha}), up to negligible
corrections of order $\mathcal{O}(H r_S)$. The scalar energy
flux~(\ref{FscalarFull}) is thus given by
$F_\text{scalar}^\text{quadrupole}/z_{\lambda,2}$, and
Eqs.~(\ref{correctAlpha}), (\ref{FGR}), (\ref{rsOnrAB}),
(\ref{correctAlphaeff}) and (\ref{zcubic}) allow us to write its
ratio to the general relativistic prediction $F_\text{GR}$ as
\begin{eqnarray}
\frac{F_\text{scalar}}{F_\text{GR}} &\approx&
\frac{2^5 \alpha_\text{BH}^2}{3^3|k_2|
\left(\Omega_\text{p} r_{AB}\right)
\left(\Omega_\text{p} r_{V23}\right)^{3/2}} \nonumber\\
&\approx& \frac{2^5 |k_2|^3}{3^4 \sqrt{3}\, k_3^2}\,
\frac{M^4}{H^3 \Omega_\text{p}}\,
\frac{1}{\left(\Omega_\text{p} r_S\right)^{5/6}}.
\label{predictionCubic1}
\end{eqnarray}
In the case of full self-acceleration, $M^4 = 3^3 k_3^2
H^4/|k_2|^3$ takes its largest possible value, and therefore also
the scalar charge (\ref{correctAlpha2}), which reaches an
$\mathcal{O}(1)$ value. This gives then the largest possible
scalar effects
\begin{equation}
\frac{F^\text{scalar}}{F^\text{GR}}
\approx
\frac{2^5}{3 \sqrt{3}}\,
\frac{H}{\Omega_\text{p}}\,
\frac{1}{\left(\Omega_\text{p} r_S\right)^{5/6}}.
\label{scalarEffects3}
\end{equation}
The presence of the very small factor $H$, of
cosmological origin, shows that the Vainshtein screening is very
efficient, and even with $\mathcal{O}(1)$ scalar charges,
experimental bounds are easily passed. The numerical
values~(\ref{Omegaprs}) and (\ref{Hrs}) indeed give
\begin{subequations}
\label{predictionsCubic2}
\begin{eqnarray}
\frac{F^\text{scalar}}{F^\text{GR}} &\approx&
6 \times 10^{-18}\text{ for LIGO/Virgo},\\
\frac{F^\text{scalar}}{F^\text{GR}} &\approx&
4 \times 10^{-11}\text{ for LISA}.
\end{eqnarray}
\end{subequations}
These predicted scalar effects are thus much smaller than the
present constraint~(\ref{ratioQuadrupole}) provided by LIGO/Virgo
data, and even the expected accuracy~(\ref{ratioLISA}) which
should be reached with LISA. They become even smaller if the
Galileon field is not responsible alone for the accelerated
expansion of the Universe, i.e., that the theory parameter $M$ is
smaller than its maximum value of order $H$, as illustrated by
Eq.~(\ref{predictionCubic1}). This can also be understood by
noting that the black-hole scalar charge~(\ref{correctAlpha}) is
proportional to $M^2$ while the Vainshtein
radius~(\ref{rV23cubic}) remains strictly the same.

In conclusion, although the time derivative of the scalar field
imposed by cosmology generates $\mathcal{O}(1)$ scalar charges
for black holes, the Vainshtein screening is so efficient that no
deviation from GR can be observed in gravitational-wave
experiments, in the quadratic plus cubic Galileon model of
Sec.~\ref{Sec4}. In other words, this full class of models passes
experimental tests. We will see below that the situation changes
drastically when considering the quintic Horndeski term.

\subsection{Simplest quintic Horndeski term}
\label{Sec6B}

In the quintic model of Sec.~\ref{Sec5} (i.e., with $k_3 = 0$),
we saw in Eq.~(\ref{smallMH}) that the scalar-field
accretion is negligible when $M/H$ is small enough. In such a
case, the relevant Vainshtein radius~(\ref{rV25-model2}) is of
the order of the size of the observable Universe, therefore
binary black holes are always deep within the screened region
where $G_5$ dominates. Since the dimensionless scalar charge
$\alpha_\text{BH}^{(c)}$, Eq.~(\ref{alphaBHc25}), is body
dependent, dipolar radiation dominates the scalar energy
flux~(\ref{FscalarFull}), and Eqs.~(\ref{FGR}), (\ref{rsOnrAB}),
(\ref{zquintic}) give
\begin{subequations}
\begin{eqnarray}
\frac{F_\text{scalar}}{F_\text{GR}} &\sim&
\frac{5 \alpha_\text{BH}^2}{96 |k_2|
\left(\Omega_\text{p} r_S\right)^{2/3}
\left(\Omega_\text{p} r_{V25}\right)^{3}}
\label{predictionQuinticLiteral}\\
&\sim&
\frac{5 |k_2|^3}{3^7 k_5^2}
\left(\frac{M}{H}\right)^8
\frac{1}{\left(H r_S\right)
\left(\Omega_\text{p} r_S\right)^{11/3}}\quad
\label{predictionQuinticLiteral2}\\
&\lesssim&
\frac{5 H}{72 \Omega_\text{p}}
\frac{1}{\left(\Omega_\text{p} r_S\right)^{8/3}},
\label{predictionQuintic1}
\end{eqnarray}
\end{subequations}
where the last inequality uses the small-accretion
bound~(\ref{smallMH}). Because of it, the tiny ratio
$H/\Omega_\text{p}$ enters again the predicted effect,
and the numerical values~(\ref{Omegaprs})-(\ref{Hrs}) give
\begin{subequations}
\label{predictionsQuintic2}
\begin{eqnarray}
\frac{F^\text{scalar}}{F^\text{GR}} &\sim&
4 \times 10^{-16}\text{ for LIGO/Virgo},
\label{predictionsQuinticLIGO}\\
\frac{F^\text{scalar}}{F^\text{GR}} &\sim&
10^{-6}\text{ for LISA}.
\label{predictionsQuinticLISA}
\end{eqnarray}
\end{subequations}
The first value is again much smaller than the present
constraint~(\ref{ratioQuadrupole}) provided by LIGO/Virgo data,
but the second is larger than our conservative LISA
accuracy~(\ref{ratioLISA}). Moreover, in the present quintic
model, we neglected the probable amplification factors similar to
$\left(\Omega_\text{p} r_{AB}\right)^{\ell-3}$ derived in
Ref.~\cite{Dar:2018dra} for the cubic case. Therefore, scalar
effects should probably be larger than our order-of-magnitude
estimate~(\ref{predictionsQuinticLISA}), especially given
$\ell=1$ for dipolar radiation. Another amplification may also
come from the larger value of $H$ to insert in
Eq.~(\ref{predictionQuintic1}), if the detected binary BH is at a
significant redshift. It is interesting to note that this
observable prediction corresponds to a model with a very small
value of $M/H$, cf.~Eq.~(\ref{smallMH}) of
Sec.~\ref{Sec5LargeAccretion}. For $10^5 m_\odot$ black holes,
this means $M/H < 6\times 10^{-5}$, and one could thus naively
think that the Galileon field has negligible influence on any
physical prediction, in the same way it can be fully forgotten in
the cosmological Friedmann equations. But because of the $1/(M
r)^2$ dependence of the $C$ term in Eq.~(\ref{coefC}), the
non-vanishing time derivative $\dot\varphi_c$ generates large
scalar charges for black holes, which do yield effects which will
be observable with the LISA interferometer.

Denoting as $\delta$
the expected bound~(\ref{ratioLISA}), LISA's consistency with GR
would imply the constraint
\begin{equation}
\frac{M}{H} <
\left(\frac{3^7 k_5^2}{5 |k_2|^3}
H r_S
\left(\Omega_\text{p} r_S\right)^{11/3}\delta\right)^{1/8}
\sim 5\times 10^{-5}.
\label{limitMH}
\end{equation}
In terms of the parameter $\alpha_T$ quantifying the speed
deviation of GWs with respect to light, Eq.~(\ref{eq:M2_aT}),
this means that, in the context of this model, LISA should be
able to constrain
\begin{equation}
|\alpha_T| < 3\times 10^{-36},
\label{limitalphaT}
\end{equation}
an improvement by 21 orders of magnitude with respect to the
present experimental limit~(\ref{eq:aT0}).

In Sec.~\ref{Sec5LargeAccretion}, we showed that for larger
values of $M/H$ than Eq.~(\ref{smallMH}),
but still consistent with a small-enough $\alpha_T$
(i.e.~ $M/H \lesssim 2\times 10^{-2}$ for $|k_2| \sim
|k_5| \sim 1$), the model predicts
large initial scalar-field accretion rates, which are eventually
quenched. The relevant Vainshtein radius is then given by
Eq.~(\ref{rV25accretion-model2}), and numerically
\begin{subequations}
\label{OmegaprV25Quintic}
\begin{eqnarray}
\Omega_\text{p} r_{V25,\text{local}} &\approx& 6 \times 10^{14}
\left(\frac{H}{M}\right)^{8/9}
\text{for LIGO/Virgo},\qquad~\\
\Omega_\text{p} r_{V25,\text{local}} &\approx& 2\times 10^{10}
\left(\frac{H}{M}\right)^{8/9}\text{for LISA}.
\end{eqnarray}
\end{subequations}
These large numbers mean again that the binaries are deep within
the screened region where $G_5$ dominates. The dimensionless
quenched scalar charge~(\ref{alpha25-low}) is still body
dependent, therefore the dipolar energy flux entering
Eq.~(\ref{FscalarFull}) dominates, and we predict again
expression~(\ref{predictionQuinticLiteral}). The difference is
that the scalar charge~(\ref{alpha25-low}) and the Vainshtein
radius~(\ref{rV25accretion-model2}) take other values, but they
finally combine to give the same
expression~(\ref{predictionQuintic1}), namely
\begin{equation}
\frac{F_\text{scalar}}{F_\text{GR}} \gtrsim
\frac{5 H}{72 \Omega_\text{p}}
\frac{1}{\left(\Omega_\text{p} r_S\right)^{8/3}},
\label{predictionQuinticLowAccretion}
\end{equation}
and numerically Eqs.~(\ref{predictionsQuintic2}). Note that this
is now a lower bound, i.e., it corresponds to the final state
where the scalar accretion rate $\Gamma_\text{acc}$ is of order
$\mathcal{O}(H)$. But if it happens that we observe a binary not
too long after the formation of the BHs, i.e., that scalar
accretion has not yet driven the scalar charges to the rather
small limit~(\ref{alpha25-low}), then we may observe larger
effects than Eqs.~(\ref{predictionsQuintic2}).

In conclusion, although the present quadratic plus quintic
Galileon model generically predicts very large scalar charges for
black holes, this also causes a large scalar-field accretion,
which makes the local value of $\dot\varphi_\text{local}$
decrease. After such an accretion, the predicted scalar effects
in binary black holes are too small to be of observational
relevance for LIGO/Virgo, but should be easily detectable with
the LISA mission. If its observations are consistent with the GR
wave templates, this means that this class of models will be
ruled out for $M/H \gtrsim 5\times 10^{-5}$,
cf.~Eq.~(\ref{limitMH}), thereby constraining $\alpha_T$ by 21
orders of magnitude tighter than the present experimental limit,
cf.~Eq.~(\ref{limitalphaT}).

\subsection{Cubic Galileon with a small quintic term}
\label{Sec6C}

Various scenarios are possible in the model of Sec.~\ref{Sec6},
depending on whether scalar accretion is negligible or
significant, and whether the BH binaries are within a $G_5$ or
$G_3$-dominated region. The clearest way to discuss them is to
plot a 2-variable diagram. It is indeed always possible to set
$|k_2| = |k_3| = 1$ by redefining $\varphi$ and $M$ in
action~(\ref{eqAction}) and functions~(\ref{G235}), and there
remains only two dimensionless parameters defining the model:
$k_5$ and $M/H$.

Figure~\ref{Fig2} illustrates the case of $10\, m_\odot$ black
holes. The region of the plane consistent with our
assumptions~(\ref{smallk5}) is the central white strip going from
the bottom-left to the top-right. In the gray triangle at the
top, one has $|k_5/k_3| (H/M)^2 < (H r_S)^2$, therefore the
quintic Galileon term is no longer dominating at small distances
$r \sim r_S$. When considering even lower values of $|k_5/k_3|
(H/M)^2$, the model ultimately tends to the (quadratic plus)
cubic Galileon discussed in Sec.~\ref{Sec6Aobs} above. In the
lower gray triangle, one has $|k_5/k_3| (H/M)^2 > H r_S$,
therefore the neglected terms in our expansions~(\ref{eqs235})
become significant, and the model ultimately tends to the
(quadratic plus) quintic Galileon discussed in Sec.~\ref{Sec6B}.

\begin{figure}[t]
\includegraphics[width=0.48\textwidth]{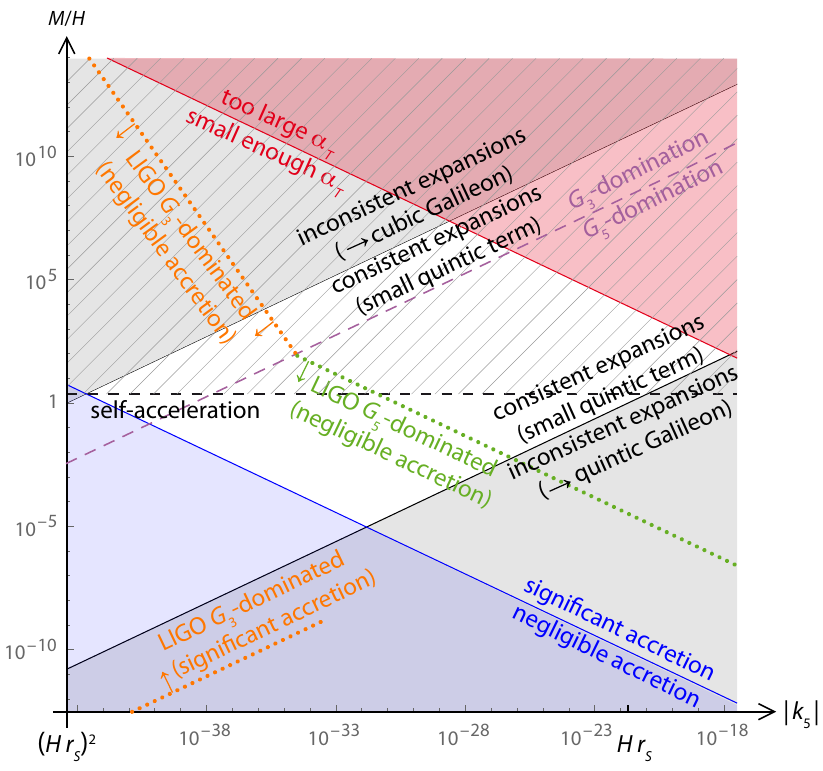}
\caption{Predictions for LIGO/Virgo in the parameter space
$(|k_5|, M/H)$, for $|k_2| = |k_3| = 1$. In the central strip
where our perturbative calculations are justified, all models
are consistent with present experimental bounds.}
\label{Fig2}
\end{figure}

The horizontal dashed line corresponds to the value of $M/H$
giving full self-acceleration. The hatched upper half-plane is
thus unphysical. The upper red triangle is forbidden by
constraint~(\ref{eq:aT0}) on the speed of GWs.

The lower blue triangle corresponds to the small values of $M/H$
such that the scalar-field accretion rate~(\ref{Gammaacc}) by the
BHs is small enough with respect to $H$. The models are also
allowed in the white region above it, but the analysis of their
predictions changes, because one must take into account the
depletion of $\dot\varphi_\text{local}$ by scalar-field
accretion, as explained in Sec.~\ref{Sec6Accretion}.

The diagonal violet dashed line separates the regions where the
binary lies within a $G_3$ or $G_5$ dominated region, implying a
different Vainshtein reduction factor~(\ref{zcubic}) or
(\ref{zquintic}).

Since the dimensionless scalar charge~(\ref{alphaBH235approx}) is
body dependent in the present model, the relevant LIGO/Virgo
observational bound is the one on dipolar radiation,
Eq.~(\ref{ratioDipole}). More explicitly, the relative scalar
contribution to the energy flux reads
\begin{equation}
\frac{F^\text{scalar}}{F^\text{GR}} \sim
\frac{5}{32 \sqrt{|k_3|}}
\,\frac{M}{\Omega_\text{p}}\,
\frac{|\alpha_\text{BH}|^{3/2}}{\left(\Omega_\text{p}
r_S\right)^{11/6}},
\label{prediction235inG3}
\end{equation}
when $G_3$ dominates the orbital physics, and
\begin{equation}
\frac{F^\text{scalar}}{F^\text{GR}} \sim
\frac{5}{48 \sqrt{2|k_5|}}
\left(\frac{M}{\Omega_\text{p}}\right)^2
\frac{|\alpha_\text{BH}|^{3/2}}{\left(\Omega_\text{p}
r_S\right)^{5/3}},
\label{prediction235inG5}
\end{equation}
when $G_5$ dominates. Note in passing that these two predictions
have the same $|\alpha_\text{BH}|^{3/2}$ dependence, in spite of
the different Vainshtein screening factors~(\ref{zcubic}) and
(\ref{zquintic}), which are respectively proportional to
$r_{V23}^{3/2}$ and $r_{V25}^3$. This comes from the fact that
$r_{V23} \propto |\alpha_\text{BH}|^{1/3}$, Eq.~(\ref{rV23}),
whereas $r_{V25} \propto |\alpha_\text{BH}|^{1/6}$,
Eq.~(\ref{rV25bis}).

The long dotted orange and green lines display the
bound~(\ref{ratioDipole}) for the models which are within the
lower blue region, i.e., with $M/H$ small enough to predict
negligible scalar-field accretion. In such a case, the scalar
charge is given by Eq.~(\ref{alphaBH235approx}). The allowed
models lie below these lines, and since the blue region is
already below them, this means that all of them pass the
LIGO/Virgo experimental tests.

On the other hand, the dotted orange line near the bottom of the
plane corresponds to the LIGO/Virgo bounds for the models lying
within the white region (where scalar accretion was initially
significant and is quenched), when $G_3$ dominates the orbital
physics and the emission of scalar waves. The corresponding
prediction is thus given by Eq.~(\ref{prediction235inG3}), but
now with the scalar charge~(\ref{alpha235-low}). The allowed
models are above this dotted line. But the corresponding models
lie in the tiny white triangle at the left of the Figure, above
the violet dashed line, and are thus already above this dotted
orange line. Therefore, once again, they all are consistent with
the LIGO/Virgo bounds.

Finally, there also exists a large white region (significant
scalar accretion followed by quenching) between the bottom blue
triangle and the violet dashed line, where $G_5$ dominates the
orbital physics. In such a case, the scalar energy flux reads
\begin{equation}
\frac{F^\text{scalar}}{F^\text{GR}}
\sim
\frac{5 H}{72 \Omega_\text{p}}\,
\frac{1}{(\Omega_\text{p} r_S)^{8/3}}
\gtrsim 4\times 10^{-16},
\label{prediction235LIGOG5}
\end{equation}
which does not depend on $|k_5|$ nor $M/H$, therefore no
corresponding line is plotted on Fig.~\ref{Fig2}. But this value
is 12 orders of magnitude smaller than the experimental
bound~(\ref{ratioDipole}), which means that all these models are
also consistent with observation. Note that
Eq.~(\ref{prediction235LIGOG5}) coincides
with~(\ref{predictionQuinticLowAccretion}) and
(\ref{predictionsQuinticLIGO}) we had found in Sec.~\ref{Sec6B}
above for the case of the quadratic plus quintic model.

In conclusion, in spite of the large scalar charges for BHs
predicted by this class of models, the Vainshtein screening and
the scalar depletion by accretion are so efficient that no
signature of scalar waves can be detected in LIGO/Virgo. Let us
recall that the scalar charge~(\ref{alphaBH235approx}) is
$\propto k_5/(H r_S)^2$ when accretion is negligible (blue region
of Fig.~\ref{Fig2}), and that its lower
bound~(\ref{alpha235-low}) after significant accretion (white
region) is still $\propto \left[|k_5| H^2/(M^4
r_S^2)\right]^{1/3}$. Because of our assumptions~(\ref{smallk5}),
which imply $|k_5| \gg |k_3| (M r_S)^2$, both these charges are
thus generically much larger than 1. The first one,
Eq.~(\ref{alphaBH235approx}), may become smaller than 1 when
$|k_5|$ is as small as our hypotheses allow \textit{and}
simultaneously $M/H \ll 1$ (bottom-left of the blue region in
Fig.~\ref{Fig2}). On the other hand, the end of accretion
limit~(\ref{alpha235-low}) is \textit{always} large within our
working interval~(\ref{smallk5}) (white region of
Fig.~\ref{Fig2}). This underlines how surprising is our
prediction of no observable signature of such BH charges in
LIGO/Virgo.
\medskip

Figure~\ref{Fig3} illustrates the case of $10^5\, m_\odot$ black
holes. The gray, hatched, red, and blue regions have the same
meaning as above, as well as the diagonal violet dashed line
separating $G_3$ and $G_5$-dominated cases. The novelty is that
LISA will be sensitive to the effect of scalar-wave radiation in
the central yellow region.

\begin{figure}[t]
\includegraphics[width=0.48\textwidth]{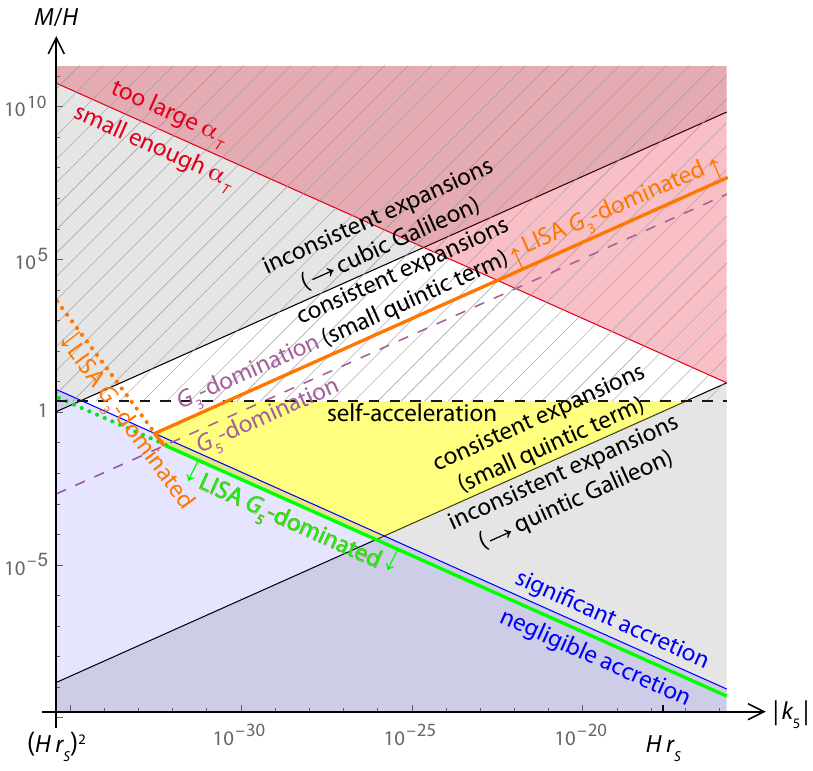}
\caption{Predictions for LISA in the parameter space $(|k_5|,
M/H)$, for $|k_2| = |k_3| = 1$. The central yellow region will be
probed, i.e., ruled out if observations are consistent with the
general relativistic GW templates.}
\label{Fig3}
\end{figure}

The orange and green lines indeed correspond to the expected LISA
constraint~(\ref{ratioLISA}) if its detections are consistent
with the general-relativistic wave templates. In the blue region
where scalar accretion is negligible, the models should lie below
the dotted orange line when the orbital physics of the binary is
dominated by $G_3$, Eq.~(\ref{prediction235inG3}), and below the
dotted or plain green lines when it is dominated by $G_5$,
Eq.~(\ref{prediction235inG5}). Since the corresponding blue
regions are already below the dotted lines, these models will be
consistent with observed data. On the other hand, the tiny strip
between the plain green and blue lines will be probed. Note that
the green line will go down when considering a less conservative
accuracy than Eq.~(\ref{ratioLISA}), i.e, a wider strip within
this blue triangle (negligible accretion) should be probed.

The white region where scalar accretion is significant is again
divided by the dashed violet line distinguishing $G_3$ and
$G_5$-dominated cases. Above this dashed violet line, i.e.,
within a white triangle at the left of the Figure, the models
below the plain orange line will be probed by LISA (they are
colored in yellow), whereas those which remain white (above the
plain orange line) will pass the tests. The remaining of the
central yellow region is dominated by $G_5$, which gives again
the literal expression~(\ref{prediction235LIGOG5}) for the scalar
energy flux. Since it does not depend on $|k_5|$ nor $M/H$, no
line is represented on Fig.~\ref{Fig3}. But its numerical value
coincides with Eq.~(\ref{predictionsQuinticLISA}) we found for
the quadratic plus quintic case of Sec.~\ref{Sec6B}, explicitly
\begin{equation}
\frac{F^\text{scalar}}{F^\text{GR}}
\sim
\frac{5 H}{72 \Omega_\text{p}}\,
\frac{1}{(\Omega_\text{p} r_S)^{8/3}}
\gtrsim 10^{-6}.
\label{prediction235LISAG5}
\end{equation}
Since it is larger than the LISA expected
accuracy~(\ref{ratioLISA}), the full central yellow region should
be probed. If LISA observations are consistent with GR, this
means that all the models within this yellow region should be
ruled out.

Note that the models along the dashed horizontal line correspond
to full self-acceleration. In such a case, Friedmann equations
give $M/H_0 = (3/|k_2|)^{3/4} \sqrt{|k_3|}$, with $H_0$ the
present value of the Hubble constant, and denoting as $\delta$
the expected LISA bound~(\ref{ratioLISA}), we get
\begin{eqnarray}
|k_5|&\lesssim&
\frac{2^7\times 3^3\sqrt{3}}{5^2 |k_2|^{3/2}}
\left(\frac{H_0}{H}\right)^2
(\Omega_\text{p} r_S)^{17/3}
(k_3\, \delta)^2
\nonumber\\
&\lesssim& 4\times 10^{-31}
\left(\frac{H_0}{H}\right)^2
\frac{k_3^2}{|k_2|^{3/2}}.
\label{k5LISAbound}
\end{eqnarray}
[It can easily be checked that $k_5 |k_2|^{3/2}/k_3^2$ is indeed
the observable notation-independent ratio entering
action~(\ref{eqAction})-(\ref{G235}).] Note that if the detected
binary is at a significant redshift, the corresponding value of
$H$ is larger than the present $H_0$, therefore the above
constraint is even stronger.

In conclusion, not only LISA should probe a significant region of
the theory plane, Fig.~\ref{Fig3}, but it should even be able to
constrain the parameter $k_5$ at the $10^{-30}$ level for
self-acceleration models. Such a tight constraint may be compared
to what Eq.~(\ref{eq:aT0}) imposes on the same self-accelerating
class of models, namely $|k_5| < 3\times 10^{-15}\,
k_3^2/|k_2|^{3/2}$. LISA will thus provide an improvement by at
least 16 orders of magnitude on the coefficient of the quintic
term. The generically large scalar charges predicted by the
present class of Horndeski theories are responsible for such a
tight bound, in spite of the Vainshtein screening and the
phenomenon of scalar accretion, which very significantly reduce
observable effects on GWs. Let us for instance mention that for
the self-accelerating model saturating
inequality~(\ref{k5LISAbound}), we still find a rather large
value of
\begin{equation}
|\alpha_\text{BH}|
\gtrsim 2 |k_2|
\left(\frac{|k_5| H^2}{3^5 k_3^2 H_0^4 r_S^2}\right)^{1/3}
\approx
15\, |k_2|^{1/2}.
\end{equation}
This means that the coupling strength of the BH to the scalar
field is still 15 times larger than to gravitons. It is obviously
even (much) larger for the other self-acceleration models at the
top of the yellow region of Fig.~\ref{Fig3}, which correspond to
values of $|k_5|$ larger by several orders of magnitude.

\section{Conclusions}
\label{Sec7}

In general relativity, there exists a so-called ``effacement
principle''~\cite{Damour:1986ny}, such that local and
large-distance physics are almost decoupled, up to tidal effects
which start manifesting on the motion of compact bodies only at
the fifth post-Newtonian order ($1/c^{10}$). Scalar-tensor
theories generically do not share this property, because the
local value of a scalar field cannot be reabsorbed in a change of
coordinates. This is the reason why the cosmological expansion of
the Universe can have an influence on local solutions, including
on black holes.

In the present paper, we have shown that the regularity of black
hole solutions imposes that they must have a scalar hair as soon
as the time derivative of the scalar field does not strictly
vanish, and even if no matter-scalar coupling is assumed in the
action. While the existence of the non-vanishing time derivatives
is driven by cosmology, this is a local effect, arising from the
requirement that a solution adopt a particular form near the
Schwarzschild radius of the black hole. The regularity of the
solution imposes a different requirement for material bodies;
nonetheless, at least the coupling to $\dot\varphi^2$ remains and
provides a contribution to the effective scalar charge.

We have shown that these charges are very large compared to
expectations in typical modified-gravity setups, despite the
extremely small scalar field backgrounds with curvatures of the
order of the Hubble rate today. For example, the quadratic plus
cubic Galileon model of Sec.~\ref{Sec4} predicts scalar charges
of order 1, meaning that black holes couple to the scalar field
with a strength similar to their coupling to gravity. This charge
causes accretion of the background scalar field onto the black
holes, but at a sufficiently small rate that a quasi-stationary
solution can be found for the whole expanding spacetime,
confirming this is a long-lived charge. Typically such a charge
would imply a change to GR predictions excluded by many orders of
magnitude. However, the Vainshtein screening of scalar effects
innate to this type of models implies that the new emission
channels are much too small to cause observable effects in
gravitational-wave experiments detecting coalescences of binary
black holes.

An even more astounding deviation from the usual setups appears
in the presence of the quintic Galileon operator. The local
consistency of the scalar field profile requires a dimensionless
black hole charge of order $(r_S)^{-2}$. Given fixed model
parameters, large enough black holes accrete slowly enough to
allow them to be considered as quasistatic solutions. On the
other hand, the charges for small black holes are large enough to
make accretion fast. Studying the evolution for such
configurations can only be done numerically and is beyond the
scope of this paper. We argue that a depletion of the charge by
the absorption of the local scalar field configuration is the
only possibility to achieve a stationary solution for a black
hole in this class of models. This charge quenching cannot remove
the charge completely, but can only reduce it to the extent that
the characteristic accretion timescale is reduced to no more than
the lifetime of the Universe. With this assumption, we show that
if LISA does not see these effects, the predicted deviation of
the speed of gravitational waves from luminal in these models
will be constrained by many additional orders of magnitude,
compared to the current bounds. We derive this phenomenology for
two models ---~a pure quadratic-quintic Galileon and a full model
containing linear $G_2, G_3$ and $G_5$ terms which we show
interpolates between the two simpler models. In the
quadratic-quintic case, for which the scalar field cannot
contribute significantly to the accelerated expansion of the
Universe, LISA should be able to improve the constraint on the GW
speed parameter $\alpha_T$ by 21 orders of magnitude. In a
self-acceleration scenario of the full model, it should improve
it by 16 orders of magnitude. Table~\ref{tab:table2}
summarizes our most important results, and points to the
corresponding equations and figure.

\begin{table}[t]
\caption{\label{tab:table2}
Summary of the most important results for the three classes
of models we consider.}
\begin{ruledtabular}
\begin{tabular}{lllll}
Scalar & Scalar & Scalar charge & Equation & LISA\\
Lagrangian & accretion & $\alpha_\text{BH}$ & number & constraints\\
\colrule
$G_2$, $G_3$ & negligible & $\mathcal{O}(1)$ &
(\ref{correctAlpha}) & no\\
\colrule
\multirow{2}{5em}{$G_2$, $G_5$} & formation & $\propto (H
r_S)^{-2}$ & (\ref{alphaBHc25}) &
\multirow{2}{4em}{Eq.~(\ref{limitMH})}\\
& quenched & $\propto (H r_S)^{-2/3}$ & (\ref{alpha25-low}) &\\
\colrule
\multirow{3}{5em}{$G_2$, $G_3$, $G_5$} & formation & $\propto k_5
(H r_S)^{-2}$ & (\ref{alphaBH235approx}) &
\multirow{3}{5em}{Fig.~\ref{Fig3} and Eq.~(\ref{k5LISAbound})} \\
& quenched & $\displaystyle \propto \left(\frac{|k_5|}{H^2
r_S^2}\right)^{1/3}$ & (\ref{alpha235-low}) &\\
\end{tabular}
\end{ruledtabular}
\end{table}

As we have pointed out, the models studied here do not exhibit
a \emph{two-horizon} problem like the one affecting the
\emph{vanilla} shift-symmetric sGB
gravity~\cite{Babichev:2024txe}, i.e., without any other
operators. Nevertheless, it is interesting that the two
situations bear some resemblance: Both the scalar charge and the
local time derivative must be adjusted uniquely for a ``good''
solution to exist, not anymore due to regularity requirements but
to those of reality and smoothness. It is not this distinction
which dooms one model and saves the others, but the fact that the
resulting value for $\dot\varphi_\text{BH}$ is unacceptably far
from $\dot\varphi_c$ in the sGB case, clashing with the assumed
cosmology. An open question is when the conditions on a solution
may become too many to be accommodated simultaneously by fixing
only two parameters, in which case the whole assumption of
stationarity breaks down, necessitating a different approach.

One may wonder to what extent our results may apply to more
general Horndeski theories of the same family, namely a
combination of $G_2(X)$, $G_3(X)$, and $G_5(X)$. For general
functions of $X$, the resulting field equation will generically
be a higher-order polynomial in $\varphi'$, leading to many more
branches of solutions and making an analytic study of the kind we
did here much more complicated, if even possible at all.
Nevertheless, there are a number of shared properties with the
models that we have considered here. Firstly, for analytic
choices of functions $G_3(X)$ and $G_5(X)$, there is always a
cosmologically-induced independent term
[Eq.~\eqref{general-source-term}] in said equation, given
purely as a function of $\dot\varphi$, acting as a source term
for $\varphi'$. For this reason we expect that black holes will
generically have hair. Secondly, not all the possible roots of
the field equation are actually relevant. Indeed, for our case of
interest there must be at least one root of the $J^0 = 0$
equation in FLRW that delivers a self-accelerating cosmological
solution, and which is given by the balance of the $G_2$ and
$G_{3,5}$-type operators (if it were possible with only $G_2$,
then the model would become effectively a much simpler
$k$-essence). Moreover, these self-accelerating solutions are in
a part of phase space for $\dot\varphi_c$ which is disconnected
{}from the Minkowski vacuum, since otherwise Minkowski would have
been the asymptotic future state. This implies that in fully
self-consistent collapse from perturbed cosmology, the solution
would not be able to transition from the timelike gradients in
cosmology to spacelike gradients of a would-be static solution
through a vacuum configurations ($X=0$) without some kind of
pathology, just as in the case presented here. This forces the
local root of the equation of motion to belong to the same vacuum
as the cosmological one. For these reasons, one may imagine that
there exists a rather large class of theories exhibiting a
similar behavior to the simpler models we have considered, where
the $G_i(X)$ functions end up acting effectively as mildly
$r$-dependent $k_i$ coefficients.

It is worth comparing the results of the present paper with
respect to known black hole solutions with a time-dependent
scalar field~(\ref{linearTime}) in scalar-tensor theories. A
particularly interesting example is the class of stealth and
self-tuning black
holes~\cite{Babichev:2013cya,Kobayashi:2014eva,Babichev:2016kdt,
Motohashi:2019sen}. These are exact solutions in Horndeski
theories and their generalizations with a scalar field of the
form~(\ref{linearTime}). For these solutions, the scalar charge
appearing in the r.h.s. of Eq.~(\ref{EqJr}) must vanish, as a
consequence of the linear time-dependence of the scalar and the
staticity of the metric~\cite{Babichev:2015rva}. One should note
that such configurations are found in theories that do not
coincide with those we considered here; in particular, for
stealth/self-tuning solutions, the $G_3$ and $G_5$ Galileon terms
should be effectively switched off~\cite{Babichev:2016kdt}. For
the cubic Galileon theory, however, non-stealth asymptotically
de~Sitter black hole solutions were found by numerical methods
in~\cite{Babichev:2016fbg}. The assumptions of staticity of the
metric and the linear time dependence of the scalar field were
imposed, implying a vanishing scalar charge of the black hole. On
the contrary, in our approach here, we assume negligible
backreaction of the scalar field in the quasi-stationary state,
thus the scalar charge needs not to be strictly zero and this is
consistent if accretion is small. Note however that
in~\cite{Babichev:2016fbg}, it was observed that numerical
integration could not be continued below some radius inside the
horizon, which might have been the consequence of missing the
branching point near the black hole, as discussed in
Sec.~\ref{Sec4C} above. This question can be resolved by studying
the dynamical formation and evolution of a black hole, which goes
beyond the scope of our paper.

Indeed there remain two large open questions for the setup we
consider, which require a separate study: \emph{(i)}~What are the
details of the evolution of the black hole and its charge in the
presence of large accretion and what is the actual end point of
this phase? Without a full numerical study, we have not been
able, for example, to estimate the duration of any such
transition to a quenched configuration, nor can we say what is
the size of the local configuration of the scalar decoupled from
cosmology. We are also working in the test-field approximation,
while it may well turn out that the end point of the quenching is
a black hole with a spacetime metric that significantly deviates
{}from Schwarzschild at small radii. However, we reiterate that
if the end point is quasi-stationary, it must have a charge, even
if the underlying solution differs from the original black hole.
\emph{(ii)}~Is the effect of non-staticity of the acoustic metric
significant? For our estimates, we relied on the only computation
of emission of scalar radiation in Vainshtein-screened regions so
far attempted, Ref.~\cite{Dar:2018dra}. Not
only is that setup purely for the cubic Galileon, as opposed to
the quintic Galileon operator of most interest here, but it is
based on the assumption of a static acoustic metric with a
standard asymptotically flat spacetime at infinity. As we have
mentioned, the presence of $\dot\varphi$ would tilt the sound
cone with respect to the light cone, e.g.~creating a
non-symmetric setup with outgoing and incoming modes propagating
at different sound speeds. The difference in the radial and
orthoradial sound speeds was important in the final result of
Ref.~\cite{Dar:2018dra} and we would presume that this will
change the details of the final answer. It is possible that the
Vainshtein screening is weakened even further than for the static
computation of Ref.~\cite{Dar:2018dra}. Moreover, the kinetic
mixing between the graviton and the scalar in the presence of the
$G_5$ term may end up complicating this static-Vainshtein picture
even further. Our computation is sufficient at the precision of
order of magnitude, but for a detailed study these open questions
must be addressed.

Understanding the answers to the above will also open another
avenue for testing these models with gravitational-wave
observations: their imprint on the ringdown phase of the merger.
The non-trivial scalar background in the vicinity of the final
charged black hole causes a significant amount of kinetic mixing
between the scalar and gravitational degrees of freedom, which is
expected to scramble the quasi-normal-mode (QNM) spectrum. The
prospect of probing dark energy in this way was considered in the
case where the shift-symmetric sGB coupling is present, acting as
the source of the scalar hair, with the other standard operators
of the Horndeski class providing the cosmology and a screening
mechanism~\cite{Noller:2019chl}. Here, the main difference is
that the sGB term is actually not needed to generate the hair,
instead being sourced by the standard Horndeski operators
themselves in the presence of the time-dependent background. We
leave the detailed study of the QNMs associated to the hairy
black-hole solutions found here for future work.

Finally, in this paper we have used a conservative method of
estimating future constraints by using information from a single
event. A population study of the type presented in
Ref.~\cite{Perkins:2020tra} should result in a stronger
constraint for the Galileon models, even without the additional
leverage of multi-band observations. However, the results of the
existing analysis cannot be applied directly, since we predict
significantly different phenomenology from the standard
dipolar-radiation case: The black hole charges and therefore the
expected radiation depend at least on the Hubble parameter at
emission as well as the black hole mass. A reanalysis of the
constraining power of future surveys is a natural follow up to
our work.

In this paper we have presented a rather unexpected result: The
cosmological boundary affects local solutions very significantly,
to the extent that we expect LISA to be able to probe
modifications of gravity relevant for cosmic acceleration. Such
complementary tests of modifications of gravity might prove key
to understand the mechanism behind dark energy, given the recent
results showing growing inconsistency of the late-time data with
the $\Lambda$CDM model~\cite{DESI:2025zgx}.

\begin{acknowledgments}
The authors would like to thank Luc Blanchet, Dražen Glavan,
Alexandre Pombo and Georg Trenkler for helpful discussions.
E.B.~acknowledges the support of ANR grant StronG
(ANR-22-CE31-0015-01). The work of L.G.T.\ was supported by the
European Union (Grant No.\ 101063210). I.S.\ is supported by the
Czech Science Foundation (GA\v{C}R) project PreCOG (Grant No.\
24-10780S).
\end{acknowledgments}

\appendix

\section{No two-horizon problem}
\label{AppendixA}

A question that we should address is whether the values of
$\alpha_\text{BH}$ and $\dot\varphi_\text{BH}$ may be conditioned
by the requirement of regularity of the scalar solutions at both
the black-hole and cosmological horizons. This has been shown to
be the case in the \emph{vanilla} shift-symmetric
scalar-Gauss-Bonnet (sGB) model, i.e., in the absence of extra
higher-dimensional operators\footnote{Such that the scalar field
equation is linear.}~\cite{Babichev:2024txe}. In that model, the
regularity conditions prevent the construction of stationary
solutions with a physically reasonable approach to homogeneity,
as $\dot\varphi_\text{BH}$ differs too strongly from
$\dot\varphi_c$. This was dubbed the two-horizon problem.
Ultimately, this is rooted in a particularity of that theory: the
$r_S^{-1}$ behavior of the sGB source term. As also stressed in
Ref.~\cite{Babichev:2024txe}, however, the mere presence of a
higher-dimensional operator is enough to resolve this tension at
the price of introducing \emph{strong} non-linearities of the
scalar field.

Let us examine these conditions in our present case of interest.
Given the shift-symmetry, we focus on the scalar quantity $X$ to
assess the regularity of the solution at each of the two
gravitational horizons, i.e., $f \to 0$. For the ans\"atze
\eqref{metric} and \eqref{linearTime}, it takes the form $X =
\dot\varphi_\text{BH}^2/f - f \varphi'^2$, and therefore a
regular $X$ implies that near a horizon $r \simeq r_h$,
\begin{equation}
\varphi'(r) \simeq \pm \frac{\dot\varphi_\text{BH}}{f}
+ \text{finite} ,
\end{equation}
where the finite parts may depend on which horizon we are looking
at. We place this form of $\varphi'(r)$ into the scalar field
equation \eqref{EqJr}, collect inverse powers of $f$ and demand
it is satisfied order by order. At the leading order $f^{-1}$,
which is the only possibly divergent one, the equation is
automatically satisfied by the above form of $\varphi'(r)$. In
other words, there are no further regularity conditions than the
one of $X$ itself. At subleading order, the constants
$\alpha_\text{BH}$ and $\dot\varphi_\text{BH}$ appear and are
balanced by the finite value $X(r_h)$ at each horizon. Eventually
higher derivatives of $\varphi'(r)$ at the horizon also show up
at higher orders. With only two horizons, these conditions are
not sufficient to fully determine $\alpha_\text{BH}$ and
$\dot\varphi_\text{BH}$, unlike the \emph{vanilla} sGB case
studied in Ref.~\cite{Babichev:2024txe}.

We conclude then that there is no two-horizon problem in the
theory we are considering here. The gravitational horizons play
no role in fixing $\alpha_\text{BH}$ and $\dot\varphi_\text{BH}$,
which may or may not be required to satisfy other conditions
elsewhere.

\section{Homogeneity for \texorpdfstring{$\dot\varphi_\text{BH}
\neq \dot\varphi_c$}{dφ(BH)/dt ≠ dφ(c)/dt}}
\label{sec:Homogeneity}

The local time derivative $\dot\varphi_\text{BH}$ of the scalar
field may differ from the cosmological background value
$\dot\varphi_c$, as studied in the context of Brans-Dicke theory
in Ref.~\cite{Glavan:2021adm}. Indeed, in order to recover a
homogeneous cosmological solution of the type
\eqref{phiBackgroundFriedmannCoords} sufficiently far away,
the only requirement is that $\varphi'(r)$ agrees with
Eq.~\eqref{phiPrimeBackground} in the large distance limit
in static coordinates ($r \to \infty$), explicitly
\begin{eqnarray}
\label{asymptotic-varphiprime}
\varphi' \simeq \frac{\dot\varphi_c}{H r}.
\end{eqnarray}
This is always satisfied by the solution to the field equation
\eqref{EqJr} in this limit.
Instead, the term linear in the ``local''-time $t$ does not
directly contribute to the linear dependence in the cosmological
time $\tau$ far away. This is an effect of the change of
coordinates of Eq.~\eqref{coord-change}. Indeed, beyond the
cosmological horizon,
\begin{eqnarray}
t = \tau - \frac{1}{2H}
\log \left[ - 1 + \left( H e^{H\tau} \rho \right)^2 \right]
\simeq - \frac{1}{H} \log \left( H \rho \right) . \qquad
\end{eqnarray}
Therefore the full $\varphi$, Eq.~\eqref{linearTime}, expressed
in cosmological coordinates in this limit ($\rho \to \infty$)
goes like
\begin{eqnarray} \label{inhomog-phi}
\varphi &\simeq& \dot\varphi_c \, \tau
+ \frac{(\dot\varphi_c - \dot\varphi_\text{BH})}{H}
\log \left( H \rho \right) .
\end{eqnarray}
One may be concerned about the fact that $\varphi \to \infty$
when $\rho \to \infty$ due to the logarithmic term. However, this
is a shift-symmetric theory and therefore the value of $\varphi$
is of no physical consequence. We should instead inspect
observable quantities such as $X$, in Friedmann coordinates
\begin{eqnarray} \label{inhomog-X}
X \simeq \dot\varphi_c^2 - e^{-2 H \tau}
\frac{(\dot\varphi_c - \dot\varphi_\text{BH})^2}{H^2 \rho^2}
\underset{\rho \to \infty}{\to} \dot\varphi_c^2 \, ,
\end{eqnarray}
which indeed approaches its expected homogeneous cosmological
limit at a (static) distance of order,
\begin{eqnarray}
r_\text{homog} \sim H^{-1} \Big|1
- \frac{\dot\varphi_\text{BH}}{\dot\varphi_c} \Big| ,
\end{eqnarray}
which should be kept not larger than $H^{-1}$. We may then allow
$\dot\varphi_\text{BH} \lesssim \dot\varphi_c$ locally to satisfy
regularity and existence conditions of the solution, without
breaking basic assumptions about cosmology nor disagreeing with
observations when the scalar drives the acceleration.
\medskip

\section{Estimate of the kinetic mixing induced by
\texorpdfstring{$G_5$}{G5}}\label{app:mixing}

To assess the importance of the kinetic mixing between the
scalar and gravitational fluctuations around the backgrounds of
interest, it will suffice to use the schematic form of this term
as it can be read from the quadratic action corresponding to our
$G_5$ term \eqref{G5}, namely
\begin{eqnarray}
Z_{5,\text{mix}} \sim \frac{k_5}{M^4} X \mathcal{R} ,
\end{eqnarray}
where $\mathcal{R}$ here represents a generic curvature quantity.
In this language, the purely scalar part discussed in
Sec.~\ref{VainshteinScreening}, Eq.~\eqref{scalar-Z5},
may be represented instead as $Z_5^{tt} \sim k_5 \mathcal{R} \nabla \nabla \varphi/M^4$.
Under the same assumptions that lead to Eqs.~\eqref{Ztt} and
\eqref{Ztt52}, the ratio of these quantities may be expressed as
\begin{eqnarray} \label{mixing-ratio-G5}
\frac{Z_{5,\text{mix}}}{Z_5^{tt}} \sim r \varphi' \sim
r^2 M^2 \left|\frac{\alpha_\text{BH}}{k_5}\right|^{1/2} ,
\end{eqnarray}
where in the second step we are considering the $G_5$ dominated
region. For the binary merger, the relevant distance at which to
evaluate this ratio is the wavelength $r \sim \Omega_P^{-1}$.
Let us define $\varepsilon_\text{mix} \equiv
\left(\Omega_\text{p} r_S\right)^2 Z_\text{mix}/Z_5^{tt}$, and
evaluate it for the cases of interest.

For the $G_2 + G_5$ model of Sec.~\ref{Sec5}, we
obtain
\begin{eqnarray}
\varepsilon_\text{mix} &\sim& \frac{M^4}{H^2} \lesssim H r_S,
\qquad \text{not quenched} \\
\varepsilon_\text{mix} &\sim& \left( \frac{H M^4 r_S^5}{|k_5|}
\right)^{1/3} \lesssim (H r_S)^{5/3}, \quad \text{quenched}\quad
\end{eqnarray}
1where we used $M \lesssim H$ to find the upper bounds, and here
$k_5 \sim \mathcal{O}(1)$ is assumed.

For the $G_2 + G_3 + G_5$ model of Sec.~\ref{Sec6}, we find
\begin{align}
\varepsilon_\text{mix} &\sim \frac{M^2 r_S}{H} \lesssim H r_S, &&
\text{not quenched} \\
\varepsilon_\text{mix} &\sim \left( \frac{M^4 r_S^3}{|\overline
k_5| H^2} \right)^{1/3} \ll H r_S, && \text{quenched}
\end{align}
where here we have replaced $k_5$ in favor of $\overline k_5$ as
defined in Eq.~\eqref{k5Bar}, and assumed in this case the
hierarchy in Eq.~\eqref{smallk5Bar}.

In conclusion, at a distance of a wavelength we may safely
neglect the kinetic mixing of the quintic operator in the cases
we have studied, and estimate the Vainshtein screening solely
from $Z_5^{tt}$ as given in Eq.~\eqref{Ztt52}.

\section{Charge implied by the cosmological branching point in
the cubic Galileon model}
\label{sec:G3-cosmo-branch}

We saw in Sec.~\ref{Sec4D} that the tiny
modification~(\ref{correctPhiDot}) of the local time derivative
$\dot\varphi_\text{BH}$ of the scalar field, with respect to its
cosmologically-imposed value $\dot\varphi_c$, suffices to make
both branching points near $r\approx \frac{3}{4} r_S$ and
$r\approx 1/(\sqrt{3}H)$ consistent with each other. It is worth
mentioning what happens when one tries to enforce
$\dot\varphi_\text{BH} = \dot\varphi_c$ strictly: One actually
finds a strong inconsistency between these two branching points.

If one assumes that the scalar charge $\alpha_\text{BH}$ takes
its value~(\ref{alphaBHc}), imposed by the inner branching point
near $r\approx \frac{3}{4} r_S$, one finds that the discriminant
$\Delta$ does \textit{not} vanish close to $r\approx
1/(\sqrt{3}H)$. It becomes very small, in the sense that
$\Delta/B^2 = \mathcal{O}(H r_S)$ for such a radius, with the
notation of Eq.~(\ref{discriminant}), but it does not admit any
real root there. However, in order for
solution~(\ref{scalarsolution}) to be close to the cosmological
background $\varphi = \dot\varphi_c \tau$, we saw in
Sec.~\ref{Sec4A} that the sign of $\pm\sqrt{\Delta}$ must be
positive at radii smaller than $1/(\sqrt{3}H)$, and negative at
larger radii. Such a change of sign would here be discontinuous,
which is impossible in absence of a singular spherical shell
source at this large distance. On the other hand, if
solution~(\ref{scalarsolution}) is continuous, then it cannot be
close to the cosmological background either at smaller or at
larger radii than $1/(\sqrt{3}H)$.

Another way to underline this inconsistency is to use again the
technique described in Sec.~\ref{Sec4B}, in order to tune
$\alpha_\text{BH}$ so that there indeed exist a double root of
$\Delta$ close to $r\approx 1/(\sqrt{3}H)$ while imposing
$\dot\varphi_\text{BH} = \dot\varphi_c$ strictly. One then finds
\begin{subequations}
\begin{eqnarray}
r_\text{root}^\text{inconsistent} &=&
\frac{1}{\sqrt{3}\, H} + \frac{3}{2}r_S
+ \mathcal{O}\left(H r_S^2 \right),
\label{rootinconsistent}\\
\alpha_\text{BH}^\text{inconsistent} &=&
\frac{\sqrt{3}\, k_2^2 M^2}{4 k_3 H} r_S
- \frac{9 k_2^2 M^2}{2 k_3} r_S^2
+ \mathcal{O}\left(H^3 r_S^3\right).
\nonumber\\
\label{alphaBHinconsistent}
\end{eqnarray}
\end{subequations}
Let us also quote this last expression when the Galileon field
$\varphi$ is assumed to be responsible alone for the cosmological
expansion:
\begin{equation}
\frac{\alpha_\text{BH}^\text{inconsistent}}{\text{sign}(k_3)
\sqrt{-k_2}} = \frac{9}{4} H r_S
- \frac{27\sqrt{3}}{2} \left(H r_S\right)^2
+ \mathcal{O}\left(H^3 r_S^3\right).
\label{alphaBHinconsistent2}
\end{equation}
We thus find that a double root of $\Delta$ would be possible at
this large distance, but for a scalar charge which is
$\mathcal{O}\left(H r_S\right)$ smaller than the one we needed at
small distances, Eq.~(\ref{alphaBHc}). Let us recall that $H r_S
\sim 10^{-22}$ (for a BH mass of $10\, m_\odot$), therefore this
is a very strong inconsistency between the two branching points.

It is thus quite surprising that the tiny
modification~(\ref{correctPhiDot}) of $\dot\varphi_\text{BH}$
sufficed to make them consistent with each other, while keeping a
linear time dependence of the scalar field~(\ref{linearTime}) in
the whole Universe. Independently of this surprise, our
result~(\ref{alphaBHinconsistent}) also illustrates the different
orders of magnitude of the scalar charges generated by branching
points close to the BH or at large distances. Other Horndeski
theories may for instance hide the inner branching point inside
the metric and sound horizons, and one should \textit{a priori}
only care about the large-distance branching points. In such a
case, one may expect to predict much smaller scalar charges than
in the present paper, and thereby negligible scalar effects in GW
detections.

\bibliography{refs}

\bibliographystyle{utphys2}

\end{document}